\documentclass[12pt]{iopart}

\usepackage{latexsym,amssymb}
\usepackage{graphics}
\usepackage{graphicx}
\usepackage{epsfig}

\begin{document}

\title{Nonequilibrium Statistical Mechanics of the
Zero-Range Process and Related Models}

\author{M. R. Evans and T. Hanney}

\address{School of Physics, University of Edinburgh,
  Mayfield Road, Edinburgh, EH9 3JZ, UK} 

\begin{abstract}
We review recent progress on the zero-range process, a model of
interacting particles which hop between the sites of a lattice with
rates that depend on the occupancy of the departure site.  We discuss
several applications which have stimulated interest in the model such
as shaken granular gases and network dynamics, also we discuss how the
model may be used as a coarse-grained description of driven
phase-separating systems.  A useful property of the zero-range process is that the
steady state has a factorised form. We show how this form enables one
to analyse in detail condensation transitions, wherein a finite
fraction of particles accumulate at a single site. We review
condensation transitions in homogeneous and heterogeneous systems and
also summarise recent progress in understanding the dynamics of
condensation. We then turn to several generalisations which also, under
certain specified conditions,
share the property of a factorised steady state.
These  include several species of particles; hop rates which depend
on both the departure and the destination sites;
continuous masses;  parallel discrete-time updating; non-conservation of particles
and sites.

\end{abstract}
\date{Jan 14, 2005}


\tableofcontents
\pagestyle{plain}

\maketitle


\section{Introduction}
\label{introduction}

The statistical mechanics of nonequilibrium systems is required to
understand the macroscopic behaviour of processes occurring throughout
physics, chemistry, biology and even sociology and economics.
Nonequilibrium phenomena are encountered whenever systems are relaxing
towards an equilibrium steady state and also whenever systems are
driven i.e.\ maintained away from equilibrium by external
forces. Systems of the driven kind, which are the main focus of this
work, cannot be described in general by equilibrium statistical
mechanics. Rather, these systems evolve to a nonequilibrium steady
state. However, though the statistical mechanics of equilibrium steady
states is well understood, analogous general principles to guide the
study of steady states far from equilibrium are only just emerging.

Considerable understanding of how microscopic interactions influence
the macroscopic properties of nonequilibrium steady states can be
gained from the study of interacting particle systems
\cite{S70,Liggett} however. These are systems defined on a lattice on
which particles hop from site to site, where the precise definition of
the stochastic particle dynamics is motivated on physical
grounds. Nonequilibrium steady states are then constructed by driving
a current of particles (a conserved quantity) through the
system. Models of this kind are known as driven diffusive systems
\cite{SZ}.

The ongoing interest in driven diffusive systems has been sustained by
the variety of the non-trivial behaviour these models exhibit. Even in
one dimension, they can undergo phase separation and phase
transitions.  These transitions may be driven by the boundary dynamics
or by defects for example, or, in translationally invariant systems,
they may be accompanied by a spontaneously broken symmetry.
Furthermore, one observes long-range (i.e.\ power-law) correlations,
not only at criticality, but generically in the steady state of driven
diffusive systems.  Notably, these properties of driven diffusive
systems in one dimension are absent from their one-dimensional
classical equilibrium counterparts.

The purpose of this article is to review recent work on a particular
driven diffusive system: the zero-range process (ZRP).  In the ZRP,
particles hop from site to site on a lattice with a hop rate which
depends, most generally, on the site from which it hops and the number
of particles at the departure site. Despite being simply stated, it
displays all of the non-trivial properties mentioned above, but with
the additional virtue that the steady state is given exactly by a
factorised form; this simple form of the steady state solution offers
an opportunity to analyse these properties exactly.

The ZRP and generalisations we discuss can be employed to investigate
fundamental aspects of nonequilibrium statistical mechanics, in that
one can address issues such as the role of conservation laws, the
range of interactions, constraints in the dynamics, and disorder, all
within the framework of an exactly solvable steady state. It has also
been widely applied as model for nonequilibrium phenomena such as
sandpile dynamics and the dynamics of avalanches, granular systems,
interface growth, polymer dynamics, various transport processes, and
glasses \cite{E00}.

The ZRP was previously reviewed in \cite{E00}. Since then this model
has stimulated considerable interest and a number of authors have
contributed to a large body of results.  Here we aim to summarise this
progress, with emphasis on developments in the `physics literature'.
In particular, a deeper understanding of condensation --- a transition
to a phase in which a single site contains a finite fraction of
particles in the system --- has emerged.  It has now become apparent
that this condensation transition appears in a number of unexpected
contexts such as wealth condensation in macroeconomies
\cite{BJJKNPZ02}, jamming in traffic \cite{E96,CSS00}, coalescence in
granular systems \cite{E99,MWL04}, gelation in networks
\cite{KRL01,BB01}. Further, we discuss the dynamics of the process by
which a condensate emerges, which have recently become better
understood.  Finally a number of generalisations of the ZRP which
retain the property of a factorised steady state have been identified
and we review these generalisations here.  We also discuss some
results that have not so far appeared in the physics literature such
as the relation between condensation in the canonical and grand
canonical ensembles (section \ref{sec:gcecond}), dynamics which
generate different ensembles (section \ref{sec:nc}) and the
Misanthrope process (section \ref{sec:Mis}).

The layout of the review is as follows.  In section \ref{definition}
we define the model and derive the steady state. In section
\ref{applications} we provide details of some of the nonequilibrium
phenomena which have recently been modelled using the ZRP. We discuss
the homogeneous ZRP (i.e.\ site-independent hop rates) in section
\ref{homogeneous} and show how the model undergoes a condensation
transition and we compare the analysis of the condensation in the
canonical and grand canonical ensembles. We then discuss the
coarsening process which determines the dynamics of the
condensation. We turn to condensation in the heterogeneous ZRP (i.e.\
site-dependent hop rates) in section \ref{sec:hetero} and show how the
mechanism of condensation and the coarsening dynamics differ from the
homogeneous ZRP. In section \ref{generalisations} we discuss several
generalisations of the ZRP, all of which are still characterised by
the exact, factorised steady state. These include generalisations to
two species of conserved particles (i.e.\ two conservation laws), to a
model in which the hop rates depend on the number of particles at both
the departure and the target sites, to continuous mass variables, to
allow arbitrary fractions of the mass at a site to hop, to a parallel
update mechanism, and to relaxing a conservation law by allowing
particle number and/or site number to fluctuate. We summarise in
section \ref{summary}.


\section{Definition and steady state} 
\label{definition}

\subsection{Definition}

The zero-range process is a model in which many indistinguishable
particles occupy sites on a lattice. Each lattice site may contain an
integer number of particles and these particles hop between
neighbouring sites with a rate that depends on the number of particles
at the site of departure.

In one dimension, the ZRP is defined on a lattice containing $L$
sites, labelled $l = 1, \ldots, L$, and we consider periodic boundary
conditions (i.e. site $L+1$ $=$ site $1$). The number of particles at
site $l$ is $n_l$, an integer greater than or equal to zero. The total
number of particles in the system is $N$ and the particle density,
$\rho$, is given by
\begin{equation}
\rho = \frac{N}{L}\;.
\label{rho}
\end{equation}
We consider totally asymmetric dynamics (although this is generalised
in section \ref{sec:arblat}) such that a particle hops from site $l$
to its nearest neighbour site to the right (site $l+1$) with rate
$u(n_l)$.

In general one can consider any lattice in any dimension, including
disordered lattices, and it turns out that one can still obtain the
exact steady state (see section \ref{sec:arblat}). For simplicity, we
will begin by considering the zero-range process in one dimension.

\subsubsection{Mapping to an asymmetric exclusion process}

\begin{figure}
\begin{center}
\includegraphics[scale=0.6]{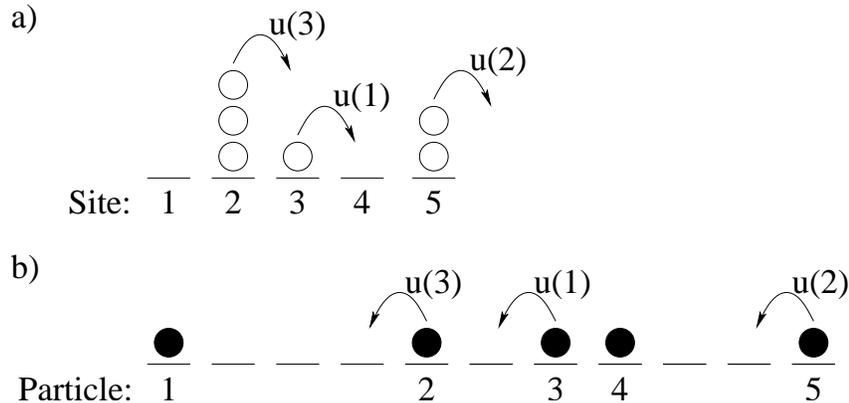}
\caption{Mapping between the zero-range process and the asymmetric
  exclusion process}
\label{mapping}
\end{center}
\end{figure}

A useful property of the zero-range process in one dimension is that
it can be mapped onto an exclusion process (i.e. a model in which 
lattice sites are either occupied by a single particle or they are
vacant). This mapping is illustrated in figure \ref{mapping}. 
The mapping is constructed by identifying a particle configuration in
the ZRP with a corresponding configuration of particles in an
exclusion model.  (The mapping is unique up to translations of the
exclusion process lattice.) To do this, one thinks of particles in the
ZRP as vacancies in the exclusion process, and sites in the ZRP as
occupied sites in the exclusion process. Thus, in figure
\ref{mapping}, site 1 in the ZRP becomes particle 1 in the exclusion
process. The next three vacancies in the exclusion process represent
the particles at site 2 in the ZRP and then site 2 itself is
represented by particle 2 in the exclusion process, and so on. In this
way one obtains an exclusion model on a lattice containing $L+N$ sites
and $L$ particles.

The exclusion process dynamics are inferred from the way in which
configurations evolve when the corresponding ZRP configurations evolve
under the ZRP dynamics: the hop rates in the ZRP, which depend on the
number of particles at the departure site, become hop rates in the
exclusion process which depend on the distance to the next particle in
front. Thus, depending on the form chosen for $u(n)$, there may be a
long-range interaction between the particles in the exclusion process.

We remark that this mapping relies on the preservation of the order of
particles under the exclusion process dynamics therefore it is only
really useful in one dimension.

\subsection{Solution of the steady state}
\label{sec:ss}
One of the most important properties of the ZRP is that its steady
state is given by a simple factorised form. This means that the steady
state probability $P(\{n_l\})$ of finding the system in a
configuration $\{n_l\} = n_1, n_2, \ldots, n_L$ is given by a product
of (scalar) factors $f(n_l)$ --- one factor for each site of the
system --- i.e. 
\begin{equation} \label{P(C)}
P(\{n_l\}) = Z_{L,N}^{-1} \prod_{l=1}^L f(n_l)\;,
\end{equation}
where $Z_{L,N}$ is a normalisation which ensures that the sum of
probabilities for all configurations containing $N$ particles is equal
to one, hence
\begin{equation}  \label{Z_L,N}
Z_{L,N} = \sum_{\{n_l\}} \prod_{l=1}^L f(n_l) \, \delta \!\left(
  \sum_{l=1}^L n_l - N \right) \;.
\end{equation}
Here, the $\delta$-function has been introduced to guarantee that we
only include those configurations containing $N$ particles in the sum.
Finally, the factors $f(n_l)$ are determined by the hop rates:
\begin{equation} \label{f(n)}
f(n) = \prod_{i=1}^{n} u(i)^{-1} \qquad {\rm for} \quad n >0\;,\qquad f(0)=1\;.
\end{equation}

We now turn to the proof of the steady state (\ref{P(C)}) to
(\ref{f(n)}). The first step is to write the steady state condition
that is satisfied by the probabilities $P(\{n_l\})$. This condition
balances the probability current due to hops into a particular
configuration with the probability current due to hops out of the same
configuration, hence
\begin{equation} \label{ME}
0 = \sum_{l=1}^{L} \left[ u(n_{l-1}\!+\!1) P(\cdots,n_{l-1}\!+\!1, n_l\!-\!1,
  \cdots) - u(n_l) P(\{n_l\})  \right] \theta(n_l)\;,
\end{equation}
where the Heaviside function $\theta(n)$ is included to emphasise that
site $l$ must be occupied for there to be associated hops out of and
into the configuration $\{n_l\}$.  The next step is to substitute the
factorised form (\ref{P(C)}) into (\ref{ME}) and look to equate each
term in the sum separately, hence
\begin{equation}
u(n_{l-1}+1) f(n_{l-1}\!+\!1) f(n_l\!-\!1) = u(n_l) f(n_{l-1}) f(n_l) \;,
\end{equation}
after cancelling common factors. This equation in turn implies that
\begin{equation}
u(n_{l-1}\!+\!1) \frac{f(n_{l-1}\!+\!1)}{f(n_{l-1})} = u(n_l)
\frac{f(n_l)}{f(n_l\!-\!1)} = {\rm constant}\;,
\end{equation}
for all values of $l$. The constant can be set equal to unity without
loss of generality, hence
\begin{equation}
f(n_l) = \frac{f(n_l\!-\!1)}{u(n_l)}\;,
\label{rec}
\end{equation}
which is readily iterated to yield (\ref{f(n)}) where we have set
$f(0)=1$, again without loss of generality. This completes the proof
of the steady state (\ref{P(C)}) to (\ref{f(n)}). 

\subsubsection{Some useful general results}
From the three equations (\ref{P(C)}) to (\ref{f(n)}) we can, in
principle, compute all the steady state properties of the ZRP. We
remark here that the steady state behaviour is determined by the form
of $f(n)$. We could, if we desired, choose {\em any} form we like for
$f(n)$ and then infer the hop rates from
\begin{equation}
u(n) = \frac{f(n\!-\!1)}{f(n)}\;,
\label{rec2}
\end{equation}
having rearranged (\ref{rec}).  The consequence of this is that any
model, with integer site variables and a conservation law, which has a
factorised steady state can be analysed within the framework of a
corresponding ZRP --- one can use the ZRP to provide a complete
account of all the possible steady state behaviour in such models.

It is important to note that $f(n)$, the single-site weight, is
distinct from $p(n)$, the probability that a given site contains $n$
particles, given by
\begin{equation} 
p(n) = f(n) \frac{Z_{L-1,N-n}}{Z_{L,N}}\;.
\label{p(n)}
\end{equation}
To obtain this equation we  fix the occupation of site one to be $n$
then sum (\ref{P(C)}) over allowed occupations of the remaining sites
subject to the constraint that the remaining number of particles is $N-n$
$$
p(n) =\sum_{n_2,\ldots,n_L} P(n,n_2\ldots n_L) \,\delta\!\left(\sum_{l=2}^L
n_l -(N-n) \right)\;.
$$
Thus the  $\delta$-function constraint in (\ref{Z_L,N}) 
is important in that it induces
correlations between sites.

It is also straightforward to obtain an expression of the mean hop
rate $\langle u(n) \rangle$
where $\langle \cdots \rangle$ denotes an average in the steady state.
 In the case of asymmetric dynamics
$\langle u(n) \rangle$ is just the particle current. In general,
\begin{eqnarray}
\langle u(n) \rangle &=& \frac{1}{Z_{L,N}} \sum_{n_1,\ldots,n_L}
u(n)\prod_{l=1}^L f(n_l)\,\delta\!\left(\sum_{l=1}^L n_l -N \right)\;,
\nonumber \\ \label{speed}
&=& \frac{Z_{L,N-1}}{Z_{L,N}}\;, 
\end{eqnarray}
where we have used (\ref{rec}). Thus the mean hop rate is independent
of location in the system so it is a conserved quantity. This result
is not necessarily obvious in the case where the dynamics are
symmetric for example, in which case $\langle u(n) \rangle$ remains
finite although the current vanishes. 

Another useful exact result is a recursion for the partition function
 $Z_{L,N}$:
\begin{eqnarray}  \label{Z_LNrec}
Z_{L,N}  &=& \sum_{n=0}^N f(n) Z_{L-1,N-n}\;,
\end{eqnarray}
which can  be easily obtained by summing (\ref{p(n)}) over $n$. This
result is usefully employed as an algorithm to be iterated on a
computer. Thus it is straightforward to compare analytic results with
numerics for $p(n)$ or $\langle u(n) \rangle$ for example.

\subsection{Generalisation to an arbitrary lattice}
\label{sec:arblat}
In this section we generalise the derivation of the steady state given
in the previous section to the case where the ZRP is defined on an
arbitrary lattice.  We use the term arbitrary lattice in a very
general sense to mean any lattice in any dimension which may include
heterogeneity in connectivity, or in the hop rates.

The steady state still factorises if the hop rate from site $l$ to
$k$, $u_{kl}(n_l)$, has the general form  
\begin{equation}
u_{kl}(n_l) = u_l(n_l) W_{kl} \;,
\end{equation}
where $u_l(n_l)$ is a site-dependent function, giving the total rate
at which a particle leaves site $l$ if $l$ is occupied by $n_l$
particles, and $W_{kl}$ is the probability that the particle hops to
site $k$. These probabilities define a stochastic matrix for a single
particle moving on a finite collection of $L$ (in this context, $L$ is
the total number of sites in the system) sites, therefore
\begin{equation} \label{stoch cond}
\sum_k W_{kl} =1\;,
\end{equation}
by conservation of probability. 

The matrix $W_{kl}$ defines an arbitrary connectivity of the
underlying lattice: if $W_{kl}$ is zero then there is no bond from
site $l$ to $k$ across which particles may hop.  Moreover, if $W_{kl}
\neq W_{lk}$ the hopping is not symmetric, thus the matrix $W_{kl}$
encodes the symmetry of the hopping dynamics.  Finally we can use
$W_{kl}$ or $u_l(n_l)$ to include heterogeneity in the hopping
dynamics.  Thus the definition of the arbitrary lattice implies these
three properties (connectivity of the lattice, symmetry and
heterogeneity).

The steady state is still given by (\ref{P(C)}) but with the factors
$f(n_l) \to f_l(n_l)$ now site-dependent. These factors are given by 
\begin{equation} \label{P(C)GEN}
f_l(n_l) = \prod_{i=1}^{n_l} \left[ \frac{s_l}{u_l(i)} \right] \qquad
  {\rm for} \quad n_l>0\;,\qquad f(0) = 1\;,
\end{equation}
where the $s_l$ are the steady state weights of a single random walker
which  moves on a lattice
with rates $W_{kl}$. These weights satisfy
\begin{equation} \label{s}
s_l = \sum_k s_k W_{lk}\;.
\end{equation}

The proof of the steady state follows the same steps as the proof
given in the previous section. The steady state condition on the
probabilities $P(\{n_l\})$ given by (\ref{ME}) is now
\begin{eqnarray}
0 = \sum_{l=1}^{L} & \left[  \sum_{k\neq l} u_k(n_{k}\!+\!1) W_{lk} 
  P(\cdots,n_{k}\!+\!1, \cdots, n_l\!-\!1,
  \cdots) \right. \nonumber \\ & \left.\phantom{\sum_{l}} - u_l(n_l)
  P(\{n_l\})  \right] \theta(n_l)\;, 
\end{eqnarray}
where we have used (\ref{stoch cond}) in the second term on the
rhs. Again, we substitute the steady state form (\ref{P(C)}) (but now
with $f(n_l)$ replaced by $f_l(n_l)$) into this equation and look to
equate each term in the sum separately, hence  
\begin{eqnarray}
u_l(n_l) f_k(n_k) f_l(n_l) = \sum_{k\neq l} u_k(n_{k}\!+\!1) W_{lk} 
  f_k(n_{k}\!+\!1) f_l(n_l\!-\!1) \;,
\end{eqnarray}
for $n_l>0$. Finally, inserting (\ref{P(C)GEN}) yields the condition
(\ref{s}). 

As an example, consider a one-dimensional chain. One can use the $W_{kl}$ to
encode partial asymmetry in the hop rates. That is, if a particle hops
to the right with rate $u_{l+1\;l}(n) = p u(n)$ and to the left with
rate $u_{l-1\;l}(n) = q u(n)$, then this corresponds to the choice
\begin{equation}
W_{l\!+\!1\;l} = p \quad {\rm and} \quad W_{l\!-\!1\;l} = q\;.
\end{equation}
With this choice, the steady state weights, $s_l$, satisfy
\begin{equation}
(p+q) s_l = p s_{l-1} + q s_{l+1}\;,
\end{equation}
with the solution $s_l =$ constant for all $l$. This constant can be
taken equal to one without loss of generality. Thus the steady state
weights (\ref{f(n)}) are unmodified by the degree of symmetry of the
hopping dynamics.  Other lattices where $s_l =$ constant are
hypercubic lattices in any dimension and the fully-connected
geometry. A consequence of this is that if the steady state factorises
for a model in one dimension it will factorise in any higher
dimension.

As an example where $s_l$ is not constant
we consider symmetric hopping on a body-centred cubic
lattice. In this case, each corner site is connected to 6 other corner
sites and 8 body-centred sites. Each body-centred site is connected to
8 corner sites. Now, assuming translational invariance, the steady
state weights, $s_c$ for corner sites and $s_b$ for body-centred
sites, satisfy
\begin{eqnarray}
s_b &=& \frac{8}{14}s_c\;.
\end{eqnarray}
Thus the steady state weights (\ref{P(C)GEN}) can be written
\begin{equation}
f_c(n_l) = \prod_{i=1}^{n_l} \left[ \frac{7}{u_l(i)} \right]\;,
\end{equation}
for corner sites and 
 \begin{equation}
f_b(n_l) = \prod_{i=1}^{n_l} \left[ \frac{4}{u_l(i)} \right]\;,
\end{equation} 
for body-centred sites. Then the steady state probabilities are
\begin{equation}
P(\{n_l\}) = Z_{L,N}^{-1} \prod_{l \in C} f_c(n_l) \prod_{l \in B} f_b(n_l)\;,
\end{equation}
where $C$ denotes the set of corner sites and $B$ denotes the set of
body-centred sites. 

More complicated $s_l$ may result from disordered hop rates,
for example a one-dimensional ZRP
with  $W_{l\!+\!1\;l} = p_l$ and  $W_{l\!-\!1\;l} = q_l$
has been studied in \cite{E96, vLK, BFL96, JSI}.

\subsection{Mathematical results}

The ZRP was has been of interest to the mathematical physics community
for a long time. It was first introduced by Spitzer \cite{S70} in 1970
and has subsequently received a lot of attention in the mathematical
literature. Of particular interest is the hydrodynamic limit. Although
this is not a limit we focus on, in this section we attempt to give a
brief overview of some of this work.

Spitzer \cite{S70} obtained the invariant measures for the ZRP for
general rates $u(n)$ for finite $L$. Subsequently these measures were
shown to be invariant even in the case of infinite $L$ \cite{A82}. The
hydrodynamic limit (i.e.\ a continuum limit yielding an equation for
the time evolution of a coarse-grained particle density field) was
obtained, for the one-dimensional symmetric ZRP in \cite{DM84}, and for the
one-dimensional asymmetric ZRP in \cite{AK84, FS94}, in which case the
hydrodynamic equation assumes the form of a nonlinear diffusion
equation. The hydrodynamic limit in higher dimensions was given in
\cite{R91, L91}.  For site-dependent rates $u_l$, without any
dependence on occupation number, the existence of invariant measures
and the hydrodynamic limit have been proven in \cite{BFL96, AFGL00}.

Steady state fluctuation properties of the homogeneous asymmetric ZRP
in one dimension were analysed by deriving macroscopic evolution
equations for a tagged particle \cite{R94} and it was shown that the
motion of such a particle is determined by the characteristic lines of
the hydrodynamic equation \cite{R95}. The size of the largest cluster,
in cases where a finite fraction of particles in the system accumulate
at a single site, was given in \cite{JMP00}, where certain conditions
on the hop rates were to given to determine whether this fraction is
less than one or equal to one (in which case, the system excluding the
cluster site contains a finite number of particles in the limit of
infinite system size).
 
This summary is far from exhaustive. In particular we have omitted
consequences of the hydrodynamic limit, such as the existence of
central limit theorems and results for large deviations. But most of
these results derive from the invariant measure in the limit of
infinite system size (i.e.$\ L\to\infty$ then $t\to\infty$) whereas
our approach is to compute the steady state on the finite lattice and
only then consider the thermodynamic limit (i.e.\ $t\to\infty$ then
$L\to \infty$).

 
\section{Applications}
\label{applications}
\subsection{Shaken granular gases}
\label{sec:sgg}
\begin{figure}
\begin{center}
\includegraphics[scale=0.5]{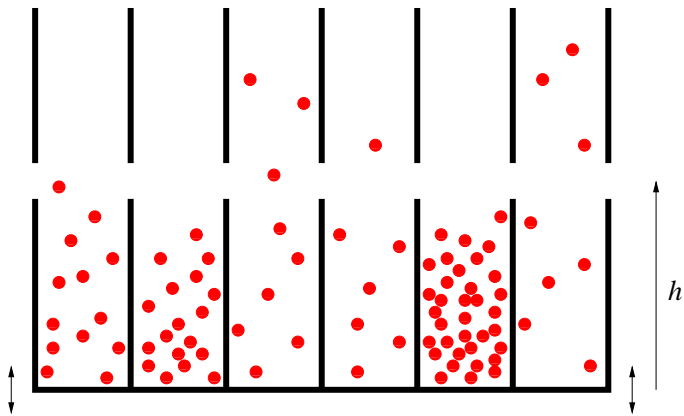}
\caption{An illustration of the vertically driven container divided into
  $L=6$ compartments which are connected by holes at height $h$. The
  system has periodic boundary conditions, so particles which leave
  the right-hand compartment to the right enter the left-hand
  compartment from the left, and {\it vice versa}.}
\label{shakensand}
\end{center}
\end{figure}

A variety of granular systems can be related to zero-range processes,
for example models of sandpile dynamics \cite{CGS93, Jain} or models of
mass transport through a series of traps \cite{BR93}. Another
example, which has received a lot of attention more recently, is
models of shaken granular gases. These models are based on experiments
in which a container is divided into $L$ equal compartments by walls,
where each wall contains a narrow horizontal slit at height $h$. The
container is then mounted on a shaker and filled with $N$ particles e.g.\
plastic balls or sand. See figure \ref{shakensand} for an illustration
of the experimental setup.
When the system is shaken vertically, the particles hop from one
compartment to another. 

The dynamics of this system clearly resemble those of the zero-range
process. To make the correspondence explicit, one has to find an
appropriate form for the ZRP hop rates $u(n)$ motivated by the kinetics of
particles within a compartment. Two forms for $u(n)$ have been
considered \cite{E99,LD02} and they have been compared, in the context
of the ZRP, in \cite{T04}.

The first of these approaches is due to Eggers \cite{E99}. An
expression for $u(n)$ is obtained by solving the equations of motion
found using the kinetic theory of vertically shaken granular
gases. For two-dimensional circular discs, the result is
\begin{equation}   \label{eggers}
u(n) = u_0 n^2 {\rm e}^{-a(n/N)^2}\;,
\end{equation}
where $u_0$ is a constant (which only sets the time scale), $n$ is the
number of particles in the compartment and $a$, given by
\begin{equation}
a=4\pi gr^2 (1-e)^2\frac{h}{A^2f^2}\left( \frac{N}{\Omega L}\right)^2\;,
\end{equation}
depends on the system parameters: $g$, the acceleration due to
gravity; $r$, the radius of the discs; $e$, the coefficient of
restitution; $A$, the amplitude of the driving; $f$, the frequency of
the driving; and $\Omega$, the width of each compartment.

For certain values of $a$ the system is found to evolve to a steady
state in which a single compartment contains most of the particles. In
the thermodynamic limit, $N\to \infty$ with $L$ fixed, this
corresponds to a phase in which the fraction of particles in a single
compartment is one. For $L=2$ this is a second order transition, from
a phase in which particles are homogeneously distributed amongst the
compartments, and is accompanied by a spontaneously broken
symmetry. The model was originally introduced for $L=2$, but has
subsequently been studied for $L=3$ \cite{WMVL01} and arbitrary $L$
\cite{MWL01,MWL02}, where it was shown the transition is first-order
for $L>2$. The dynamics of the coarsening leading to the condensed
phase was studied in \cite{MWL04}. Experiments and simulation indicate
that the system evolves to its steady state in the condensed phase in
two stages: firstly, most of the particles cluster in a few
compartments, then these clusters disappear one by one until only a
single cluster remains. The mean cluster size is found to grow like
$[{\rm ln}(t)]^{1/2}$. The appropriate form of the hop rates for a
bidisperse granular gas was derived in \cite{MMWL02}; this model
corresponds to a ZRP with two different types of particles, a
generalisation of the ZRP we discuss in section \ref{generalisations}.
Finally, an asymmetric case where a current of particles is supported
due to holes connecting compartments at different heights has been
studied \cite{MRWL}. In this case the ZRP description does not yield a
factorised steady state since the hop rates do not fall into the class
discussed in section \ref{sec:arblat} but the assumption of
factorisation provides a mean-field-type approximation.

The second approach is due to Lipowski and Droz \cite{LD02}, who
propose a simpler model than that of Eggers which still captures the
essence of the phenomena. They consider a hop rate of the form
\begin{equation} \label{LD}
u(n) = \frac{n}{N} {\rm exp}\left( -\frac{1}{T_0 + \Delta(1-n/N)}\right)\;,
\end{equation}
where $T_0$ and $\Delta$ are positive constants, which depend on
system parameters. The basis of Eggers' derivation of (\ref{eggers})
was that the effective temperature of a granular system decreases when
the particle density increases. The form (\ref{LD}) is the simplest
hop rate which reproduces this fact, due to the linear $n$-dependence
in the denominator of the exponential. The choice (\ref{LD}) then
describes dynamics under which a particle is selected at random and
moved to a randomly chosen neighbouring compartment with probability
${\rm exp}( -1/[T_0 + \Delta(1-n/N)])$.

As in Eggers' model, this model undergoes a transition between a
homogeneous phase and a phase in which most of the particles occupy a
single compartment. Again, the transition is second-order for $L=2$
\cite{LD02,SPL03} and first-order for $L>2$
\cite{CDL02,SPNL04}. During the coarsening stage of the dynamics the
mean cluster size is found to grow like ${\rm ln}(t)$
\cite{LD02,T04,SPL03,CDL02,SPNL04}.

\subsection{Networks}
\label{sec:net}
A network is defined as a set of nodes interconnected by links
---mathematically a network is simply a graph and the links, which may
be directed or undirected, are the edges of the graph.  The interest
within the statistical physics community into networks has been with
regard to the statistics of the connectivity properties.  For example,
the degree of a node is the number of links attached to it and the
degree probability distribution may exhibit a `fat tail' or even a
power-law asymptotic behaviour.  Simple models to generate dynamically
such networks and statistics have been introduced and studied
extensively (for recent reviews see \cite{AB02,DM03}).

\subsubsection{Growing and rewiring  networks}

The dynamics of growing networks is introduced through stochastic
rules for attaching links to nodes.  Basically there are two types of
dynamical networks: growing networks where links and nodes are added
to the networks \cite{BA99}  and `re-wiring' or `equilibrium'
networks where the number of nodes and links is fixed but links
re-attach stochastically from one node to another \cite{WS98,DMS03}

In growing networks, power-law degree distributions may be generated
through preferential attachment.  The idea is to add a new node to the
network at each time step and to attach this node to an existing node
selected with probability proportional to its degree.  It has been
shown that the linear dependence of attachment rate on degree gives
rise to a power-law degree distribution \cite{BA99}.  If the attachment
rate grows more slowly than linearly in the degree of the node one can
generate stretched exponential degree distributions; if the attachment
rate grows faster than linearly in the degree of the node, a
condensate node is generated which is connected to nearly all other
nodes \cite{KR01,KRL01}.

Generally, condensation (also referred to as gelation in the network
literature) occurs when a node captures a finite fraction of the total
number of links.  As well as condensation induced by nonlinear
preferential attachment one can have condensation induced by
heterogeneity \cite{BB01}. Here nodes have an intrinsic fitness and
the rate of attachment is proportional to this. The two types of
condensation are reminiscent of the two types of condensation seen in
the ZRP to be discussed in sections \ref{sec:homcon} and
\ref{sec:hetero}. The connection is explicit in the case of rewiring
networks which we review in more detail in section
\ref{sec:rewire}. In these networks the nodes correspond to sites of
the lattice and the dynamics of links becomes equivalent to the
dynamics of the particles in a generalised ZRP on a fully-connected
geometry.

\subsubsection{Directed edges and two-species ZRP}
One can also consider networks where the edges are directed. Then each node
is characterised by its in-degree and out-degree which are the
number of edges pointing into the node and the number of edges
pointing out of the node respectively.  Rewiring dynamics consists of
a rate for rewiring the outgoing end of an edge and a rate for
rewiring the ingoing end of an edge \cite{DMS03} and these rates most
generally depend on the in- and out-degree of both the
departure and destination nodes.

If we think of ingoing edges and the outgoing edges as two species of
particles we have  a generalisation of the two-species ZRP discussed in
section \ref{sec:2szrp}.  Again the model is defined on a fully-connected
geometry and one obtains a condition for factorisation similar in form to
the factorisation condition for the two-species ZRP \cite{DMS03}. It should be
possible to analyse condensation transitions within this framework

\subsection{Coarse-grained descriptions}
So far we have presented applications of the ZRP which essentially
involve mapping various models on to the ZRP at the level of exchange
of particles between sites of the system.  More generally, however,
one may think of the sites of the ZRP as representing domains of some
driven system---this is a most natural picture within the exclusion
process interpretation of the ZRP (figure \ref{mapping}).  The domains
may have some internal structure, for example further degrees of
freedom, but this all integrated out, and one is left with an
effective dynamics of exchange of length between domains given by
$u(n)$.

\subsubsection{Bus route model}
An early example of the use of the ZRP as an effective coarse-grained
description of more complicated microscopic dynamics is the `bus route
model' \cite{OEC98}.  The model is defined on a $1d$ lattice. Each
site (bus-stop) is either empty, contains a bus (a conserved particle)
or contains a passenger (non-conserved quantity). The dynamical
processes are that passengers arrive at an empty site with rate
$\lambda$; a bus moves forward to the next stop with rate 1 if that
stop is empty; if the next stop contains passengers the bus moves
forward with rate $\beta$ and removes the passengers.  The bus route
model can be related to the ZRP by an approximation of a mean-field
nature in which we integrate out the non-conserved quantity
(passengers). The bus-route dynamics can be thought of as
an exclusion process   which is  mapped onto  ZRP
dynamics as in figure \ref{mapping}:
the buses correspond to the
sites of the ZRP and the number of bus stops between a bus and the
next bus along the route, corresponds to the number of particles at
the site of the ZRP.
The idea is that
the hop rate of the buses becomes a function of the distance
to the next bus ahead. Now,
the mean time elapsed since a
bus-stop was last visited is given by
$n/v$ where $n$ is the distance to the next bus ahead
and $v$ is the steady-state speed.  Therefore the mean-field
probability that the site next to a bus is not occupied by a passenger
is $\exp(-\lambda n/v)$. From this probability an effective hop
rate for a bus into a gap of size $n$ is obtained by averaging the two
possible hop rates $1,\beta$:
\begin{equation}
u(n)=\beta+(1-\beta) \exp(-\lambda n/v)\;.
\end{equation}
This example serves to illustrate how local dynamics may generate an
effective ZRP dynamics defined by $u(n)$.

The phenomenon of condensation, wherein a single site of the ZRP
contains a finite fraction of the density (to be discussed in
\ref{sec:homcon}), would correspond to a
finite fraction of the bus stops being between two buses i.e.  there
is a jam of buses in the steady state and all the buses arrive at a
bus stop at once!  However it turns out that since $u(n)$ decays
exponentially the condition for a strict phase transition in the
thermodynamic limit is not met, unless we take the passenger arrival
rate $\lambda \to 0$.  However on any {\em finite} system for
$\lambda$ sufficiently small, an apparent condensation will be seen.

In \cite{OEC98}  a dual model to the bus-route model was used to
model clogging in pipes. Also a model of ant trails introduced in
\cite{CGNS02} may be mapped exactly onto the bus route model.

\subsubsection{Phase separation in one-dimensional systems}
\label{sec:ps}
Recently, using the ZRP as the effective description has allowed
insight into the long-standing question of phase separation in
one-dimensional driven systems \cite{KLMST02}.  Within this
description phase separation is manifested by the emergence of one
large domain and this corresponds to the phenomenon of condensation in
the ZRP.  The criterion for phase separation corresponds to the
criterion for condensation within the ZRP \cite{KLMST02,ELMM04}.

The idea is that domains of particles on a one-dimensional lattice can
be represented by the sites of a ZRP and the dynamics by which domains
exchange length can be described by $u(n)$.  Thus, in this context,
$u(n)$ corresponds to a conserved current flowing out of a domain of
size $n$ and the assumption is that domains are uncorrelated in that
the current only depends on the length of a domain. The approach
reduces a many-domain problem to the properties of a single
domain; the key step is the identification of the effective current
$u(n)$.

As we shall see in section \ref{sec:u(n)cond} the criterion to have
condensation is that either the hop rates $u(n)$ in the ZRP should
decrease to 0 as $n$ increases or else $u(n)$ should decay more slowly
to an asymptotic value $\beta$ than $\beta(1+2/n)$. Thus, the
criterion for phase separation to occur is that the current of
particles out of a domain of size $n$ should either decay to zero with
domain size or decay to a non zero value more slowly than
$\beta(1+2/n)$.

In the development of this approach particular attention has focussed
on a class of exclusion models with positive particles, negative
particles and vacancies \cite{SHZ}.  The model we consider is defined
on a one-dimensional ring of $L$ sites. Each site $i$ is associated
with a `spin' variable $s_i$. A site can either be vacant ($s_i=0$) or
occupied by a positive ($s_i=+1$) or a negative ($s_i=-1$)
particle. Particles are subject to hard-core repulsion and a
nearest-neighbour `ferromagnetic' interaction, defined by the
potential
\begin{equation}
V = - \frac{\epsilon}{4} \sum_i s_i s_{i+1}\;.
\end{equation}
Here $0 \leq \epsilon < 1$ is the interaction strength, and the
summation runs over all lattice sites. The model evolves according
to the nearest-neighbour exchange rates
\begin{equation}
\label{eq:rates} 
+- 
\mathop{\longrightarrow}^{(1+{\rm\Delta} V)}
-+\qquad
-+ 
\mathop{\longrightarrow}_{q(1+{\rm\Delta} V)}
+-
\qquad +0 \mathop{\longrightarrow}^\alpha 0+ \qquad 0-
\mathop{\longrightarrow}^\alpha -0\;,
\end{equation}
where ${\rm\Delta} V$ is the difference in the potential $V$ between
the initial and final states.  One can think of this as a
generalisation of the Katz-Lebowitz-Spohn driven lattice gas (KLS
model) to be discussed in section \ref{sec:KLS}.

\begin{figure}
\begin{center}
\includegraphics[scale=0.5]{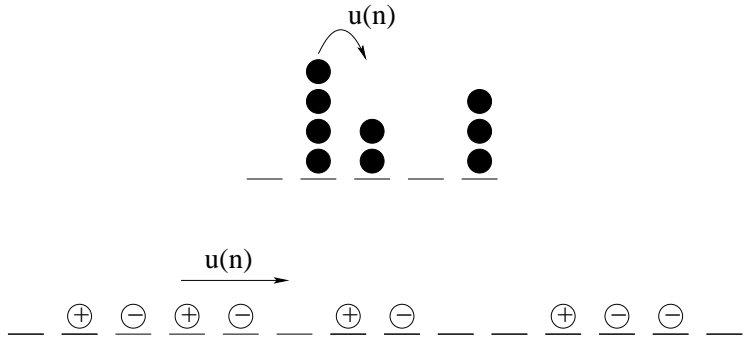}
\caption{A microscopic configuration of a two-species driven model
(bottom) and its corresponding configuration in the ZRP (top).}
\label{fig:cartoon}
\end{center}
\end{figure}

The sites of the exclusion process correspond to the vacancies in this
model and the domains to the consecutive sequences of positive and
negative particles as illustrated in figure \ref{fig:cartoon}.  Note
that although a domain comprises both positive and negative particles,
the effective description is in terms of a one-species ZRP.  In the
case $q>1$ (where positive particles drift preferentially to the left
of negative particles) the steady state comprises three pure regions
of vacancies, positive particles and negative particles.  This
corresponds to strong condensation where one site of the ZRP holds a
fraction tending to one of the particles. The current of particles is
exponentially small in the system size \cite{Blythe00} and $u(n)$
decays exponentially to zero.  A more subtle question is whether
condensation and phase separation can occur for $q <1$.

In the case $\epsilon=0$ the model reduces to that studied in
\cite{DJLS,AHR} and an exact solution is available via a matrix
product ansatz \cite{DEHP93}.  The mapping of the steady state of the
exclusion process to the steady state of a ZRP becomes exact and
allows an exact identification of the effective current $u(n)$.  In
the case of equal densities of positive and negative particles the
function $u(n)$ is precisely equal to the current flowing in the
steady state of an exclusion process of length $n$ with open boundary
conditions \cite{KLMST02}.  This gives a simple recipe for determining
$u(n)$ and provides a numerical test for phase separation, even when
the open system cannot be solved exactly \cite{KLMST02}.  In the case
of nonequal densities, the current is given by an exclusion process of
length $n$ with periodic boundary conditions but with non conservation
of the charge of particles \cite{ELMM04}.  In either case the current
decays with domain size as $u(n) = u(\infty)(1+ 3/(2n))$ which does
not satisfy the condensation criterion.  However one may have very
strong finite size effects leading to sharp crossover effects which
may give the impression of a phase transition on finite systems
\cite{RSS,KLMT02}.

More generally, when $\epsilon \neq 0$, and the ferromagnetic
interaction is switched on, no exact solution is available.  However
using the identification of $u(n)$ suggested by the $\epsilon =0$ case
predicts that phase separation (for $q <1$) should occur when
$\epsilon$ is sufficiently large \cite{KLMSW03}.  This effect is
enhanced when the densities of positive and negative particles are
unequal \cite{ELMM04}.  Although the mapping to a ZRP is not expected
to be exact when $\epsilon \neq 0$, its predictions have proven
accurate so far.


\section{Homogeneous system}
\label{homogeneous}
We now return to the ZRP defined in section \ref{definition}.  In this
section we study the homogeneous system in which hopping rates $u(n)$
and therefore steady-state weights $f(n)$ are site independent (see
section \ref{sec:ss}).  We will focus on the phenomenon of
condensation.  As we shall see the condition for this to occur depends
on the asymptotic behaviour of $f(n)$ and therefore $u(n)$.  We also
compare the analysis of condensation in the canonical and grand
canonical ensembles.

As we have seen in the introduction and section \ref{applications}
condensation is observed in a variety of physical contexts.  The
factorisation property allows us to analyse exactly the condensation
mechanism within the ZRP.  The first analysis of this kind on a
homogeneous system was carried out in \cite{BBJ} where the grand
canonical partition function of a `balls-in-boxes' model is analysed.
It should be noted that condensation also occurs in other systems that
cannot be solved exactly and a factorised steady state becomes an
approximation within which to study condensation \cite{MKB,LMZ}.

Ideally one would wish to demonstrate condensation by analysing the
distribution of the number of particles $p(n)$ given by (\ref{p(n)}).
This expression involves $Z_{L,N}$, defined by (\ref{Z_L,N}), which is
the equivalent of the canonical partition function since the number of
particles is held fixed. However, to understand the condensation
phenomenon, it is simplest to work within the grand canonical ensemble
where the particle number fluctuates.

\subsection{Grand canonical ensemble}
\label{sec:gce}
We define the grand canonical partition function as
\begin{equation}
{\mathcal Z}_L(z) = \sum_{n=0}^\infty z^n Z_{L,n}\;,
\label{Zgc}
\end{equation}
where $z$ is the fugacity which is chosen to fix the density $\rho$
through
\begin{equation}
\rho = \frac{z}{L} \frac{\partial \ln {\cal Z}_L(z)}{\partial z}\;.
\label{rhogc}
\end{equation}
Now,  using (\ref{Z_L,N}), we have
\begin{equation}
{\cal Z}_L(z) =  
\sum_{\{ m_l =0\}}^{\infty} z^{\sum_l m_l}
 \prod_{l=1}^{L} f(m_l) \; 
= \left[ F(z)\right]^L
\label{Zgc2}
\end{equation}
where 
\begin{equation}
F(z) = 
 \sum_{m=0}^{\infty} z^{m} \ f(m)\; 
\label{Fdef}
\end{equation}
and
\begin{equation}
\rho = z \frac{F'(z)}{F(z)}
\;.
\label{rhocon}
\end{equation}
One views this condition either as defining explicitly $\rho(z)$, or
as defining implicitly $z(\rho)$.

The distribution of the number of particles at a given site becomes
\begin{equation} \label{p(n)gc}
p(n) = z^n f(n) \frac{{\cal Z}_{L-1}}{{\cal Z}_{L}}= \frac{z^n f(n)}{F(z)}\
\end{equation}
in the grand canonical ensemble.  We can then calculate the mean hop
rate $\langle u(n) \rangle= \sum_n p(n) u(n)$ as
\begin{equation}
\langle u(n) \rangle = z\;,
\end{equation}
endowing the fugacity with a physical interpretation.  We emphasise
that here the angle brackets denote a steady state average within the
grand canonical ensemble. This is to be compared with the expression
of the mean hop rate in the canonical ensemble, given by (\ref{speed}).

\subsection{Condensation}
\label{sec:homcon}
It is easy to show, by taking the derivative with respect to $z$, that
the right hand side of (\ref{rhocon}) is an increasing function of
$z$. Also let the radius of convergence of $F(z)$, defined in
(\ref{Fdef}), be $z= \beta$. Thus increasing the value of $z$
corresponds to increasing values of the density $\rho(z)$ until $z$
takes its maximum value $\beta$ which corresponds to the maximal value
of the density $\rho_c$
\begin{equation}
\rho_c = \beta \frac{F'(\beta)}{F(\beta)}\;.
\label{rhoc}
\end{equation}
If $\rho_c$ is infinite then for any finite density one can find a
solution of $(\ref{rhocon})$ for $z$. On the other hand, if $\rho_c$
is finite then for $\rho > \rho_c$ one can no longer satisfy
$(\ref{rhocon})$ and we have condensation.

Let us be specific and consider $f(n)$ with large $n$ asymptotic form
\begin{equation}
f(n) \sim \frac{A}{\beta^n n^b}\;.
\label{f(m)a}
\end{equation}
Then $z= \beta$ is clearly the radius of convergence of (\ref{Fdef})
and whether $\rho_c$ is finite or not is controlled by $F'(\beta)$
which converges for $b >2$. Thus for $b>2$ we will have condensation.

To understand what is happening we look at the single-site
distribution $p(n)$ (\ref{p(n)}) for the number of particles.  For
$n\gg 1$ this becomes, using the asymptotic form (\ref{f(m)a}),
\begin{equation}
p(n) \simeq \frac{(z/\beta)^n}{F(z)} \frac{A}{n^b}\;.
\label{p(n)a}
\end{equation}
Consider first $b\leq 2$ in which case $\rho_c = \infty$.  The
distribution (\ref{p(n)a}) is a power law $\sim 1/n^b$ with an
exponential prefactor cutting off the distribution at $n \sim
1/\ln(z/\beta)$. Increasing values of $z$ imply that the distribution
is cut off at larger values of $n$ and thus correspond to increasing
values of the density. If $b \leq 2$ one can obtain any desired
density $\rho = \sum_n n p(n)$ by choosing $z$ to cut off the power
law at suitable $n$.

For $b > 2$, on the other hand, the power law $1/n^b$ has a finite
mean. Thus letting the cut-off tend to infinity (i.e. $z\to \beta$)
only corresponds to a finite density. Therefore for $\rho >\rho_c$ one
requires an extra piece to the distribution which represents the
condensate. The shortfall in the density, $\rho-\rho_c$, will be made
up by the presence of this piece.
\begin{figure}
\begin{center}
\includegraphics[scale=0.5,angle=270]{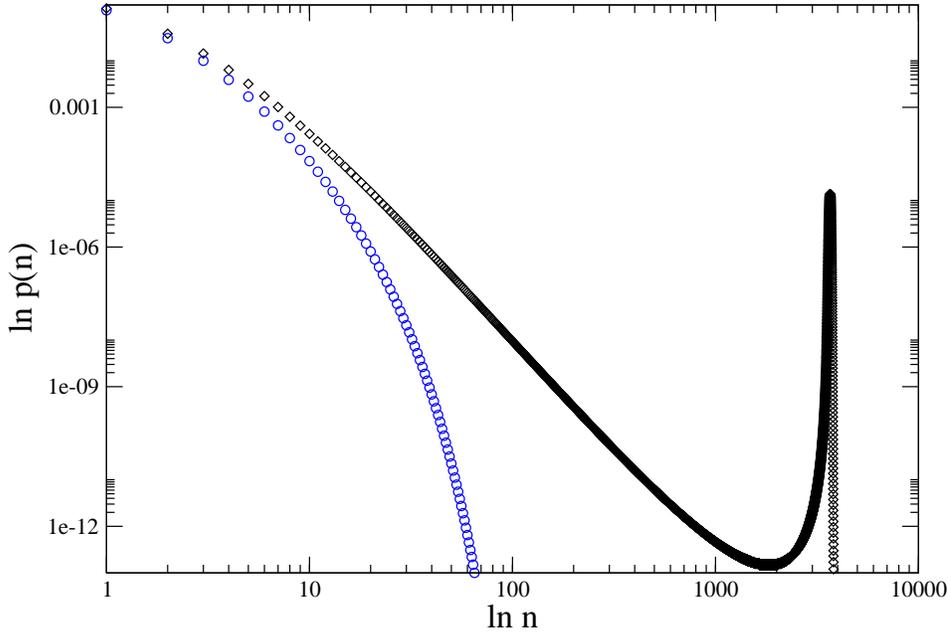}
\caption{ln-ln plot of the single-site distribution $p(n)$ v.\ particle
  number $n$. The data are obtained by iterating the recursion
  relation (\ref{Z_LNrec}) for $Z_{L,N}$ for $L=1000$ and $b=5$. The
  circles represent $\rho=1/4$ where the system is in the fluid phase;
  the diamonds represent $\rho=4$ where the system is in the condensed
  phase.}
\label{can_dist}
\end{center}
\end{figure}

To confirm that the condensate corresponds to a single site we
consider its canonical weight: if there is a single condensate site
the weight will be $Lf(N-N_c) = {\mathcal O}(L^{1-b})$, whereas if the
condensate is spread over two sites the weight will be $ {L \choose 2}
\sum_n' f(N- N_c-n)f(n) = {\mathcal O}(L^{3-2b})$, where the prime
indicates a sum over terms $n= {\mathcal O} (N)$.  The factors $L$ and
$ {L \choose 2}$ come from the number of ways of choosing the
condensate site(s).  Now the condition for condensation, $b>2$,
implies that $1-b > 3-2b$, therefore in the limit $L\to \infty$
configurations with a single condensate are dominant over
configurations in which the condensate is spread over two (or more)
sites.  Note that in the condensed phase we have used canonical
weights, because in this phase the equivalence between ensembles
breaks down as we discuss in the next subsection. A rigorous proof
that the condensate corresponds to a single site is given in
\cite{GSS03}.

\subsection{Relation between canonical and grand canonical partition
functions}

We now return to the calculation of the canonical partition function
and the question of equivalence between canonical and grand canonical
ensembles.

 We may write $Z_{L,N}$  using Cauchy's theorem as
\begin{equation}
Z_{L,N} = \oint \, \frac{ds}{2\pi i} \ s^{-(N+1)}\ 
{\cal Z}_L(s)
 = \oint \, \frac{ds}{2\pi i} \ s^{-(N+1)}\ \left[ F(s)\right]^L\; ,
\label{Zint}
\end{equation}
where the integral is around a closed contour about the origin
in the complex $s$ plane.
For large $N,L$ (\ref{Zint}) is dominated by the saddle point of the
integral which we denote by $s=z$.  Defining
\begin{equation}
\phi(s)  = -\rho \ln s + \ln[ F(s) ]
\end{equation}
the saddle point $z$ is given by 
$\phi'(z) =0$  and this  recovers precisely
(\ref{rhocon}).
Moreover
\begin{eqnarray}
Z_{L,N} &\simeq& \frac{1}{(2\pi L)^{1/2}} \frac{1}{\left| \phi''(z)\right|^{1/2}}\frac{{\rm e}^{L\phi(z)}}{z}\\
&=& \frac{1}{(2\pi L)^{1/2}} \frac{1}{\left| \phi''(z)\right|^{1/2}}
\frac{ {\cal Z}_L(z) }{z^{N+1}}
\label{Zgr}
\end{eqnarray}
Thus when one has a valid saddle point the canonical and grand
canonical ensembles are equivalent \cite{Huang}.  However, for a valid
saddle point the value of $z$ must be within the radius of convergence
of $F(z)$ since the integral contour cannot be extended to enclose the
singularity at $s=\beta$. Thus for $\rho >\rho_c$ the asymptotics of
the canonical partition function must take on an alternative form,
signalling the phase transition.  Only recently has a detailed
analysis of this form and the condensation mechanism been carried out
\cite{EMZ05}; we give a brief discussion here and refer the reader to
\cite{EMZ05}. Also note that the equivalence of ensembles has been
studied in \cite{GSS03}. There it was shown that for $\rho >\rho_c$
the fluid part of the canonical distribution coincided with the
critical grand canonical distribution.

\subsubsection{Condensation within  canonical ensemble}
As $\rho \nearrow {\rho_c}$ the saddle point $z$ approaches
a branch cut singularity at $z=\beta$. For $\rho \geq {\rho_c}$ 
there is no valid saddle point, but the integral (\ref{Zint})
is still  dominated by the region $s \simeq \beta$. 
Thus in the critical region and in the whole of the condensed phase
one can expand about $s=\beta$.
For $f(n)$ of the form (\ref{f(m)a}) we
let $s=\beta(1-u)$ and expand $F(s)$ for $u$ small  as
\begin{equation}
F(\beta(1-u))
= \sum_{k=0}^{r-1} \frac{  (-\beta u)^k F^{(k)}}{k!}
+ B u^{b-1} + \ldots
\label{FB1u}
\end{equation}
where $r$ is the integer part of $b$
and $F^{(k)} = \frac{d^k}{d s^k}\left.  F(s)\right|_{s=\beta}$
(see e.g. \cite{BB96}).
The  second term in (\ref{FB1u}) is the leading singular part and 
the coefficient $B$  is determined through
\begin{equation}
B= \lim_{u\to 0} \frac{u^{r+1-b}}{(b-1)\cdots(b-r)} \frac{d^r F}{d u^r}
\end{equation}
and it can be shown that
\begin{equation}
B= A \Gamma(1-b)\;,
\end{equation}
where $\Gamma(x)$ is the usual gamma function. 

In the condensed phase the contour integral in (\ref{Zint}) can be deformed
to run along the imaginary axis in the complex $u$ plane. Then
the  asymptotic behaviour is controlled by 
the nonanalytic term $A \Gamma(1-b)u^{b-1}$ and one obtains
\begin{equation}
Z_{L,N> N_c} 
\sim \frac{F(\beta)^L}{\beta^N} \int_{-i\infty}^{i\infty}
\frac{du}{2\pi i } 
\exp L \left[  u(\rho-\rho_c) + \ldots+ u^{b-1}
\frac{A \Gamma(1-b)}{F(\beta)} + \ldots\right]\;.
\end{equation}
The asymptotics of this integral can be evaluated by wrapping the contour
around the negative real axis and one finds that the leading
contribution is
\begin{equation}
Z_{L,N> N_c} 
\simeq \frac{F(\beta)^{L-1}}{\beta^N}
\frac{A}{(\rho-\rho_c)^bL^{b-1}}
\label{Zc}\qquad \mbox{all noninteger}\quad b>2\;.
\end{equation}
Thus we have a distinct supercritical form for the canonical partition
function (\ref{Zc}) to be compared with the grand canonical form
(\ref{Zgr}) which holds in the subcritical regime. These forms can be
used to analyse the piece of $p(n)$ which represents the condensate
and one finds distinct universal forms for $2<b<3$ and $b>3$
\cite{EMZ05}.

\subsubsection{Condensation within grand canonical ensemble}
\label{sec:gcecond}
In contrast, an analysis of the condensate within the grand canonical
ensemble requires allowing $z> \beta$, thus one has to impose cut-offs
on the sums \cite{PS} and modify (\ref{Zgc2}) to
\begin{equation}
{\cal Z}_L(z) 
= \left[ F_N(z) \right]^L\quad\mbox{where}\quad
F_N(z)= \sum_{n=0}^N f(n) z^n \;.
\label{Zgc3}
\end{equation}
Here the cut-off $N$ ensures that no site contains more than $N$, the
total number of particles in the canonical ensemble, but where the
number of particles still fluctuates.  Then to satisfy (\ref{rhoc})
for $\rho>\rho_c$ one takes $z= \beta(1+ \eta)$ with $\eta$ small and
chooses $\eta$ so that $\beta F'(\beta) <z F_N'(z) ={\mathcal O}(1)$.
The result is \cite{AEM04}
\begin{equation}
z= \beta(1 + \eta)\;,\qquad \eta= (b-2) \frac{\ln N}{N} + \frac{\ln
  \ln N}{N} + {\mathcal O}\left( \frac{1}{N}\right)\;, 
\end{equation}
where the coefficient of the ${\mathcal O}(1/N)$ term has to be chosen
to ensure that
(\ref{rhoc}) is satisfied.  This result implies that the mean hopping
rate, $\langle u \rangle= z$ in the grand canonical ensemble,
overshoots its thermodynamic limit value $\beta$ on a large but finite
system \cite{AEM04}.

\begin{figure}
\begin{center}
\includegraphics[scale=0.5,angle=270]{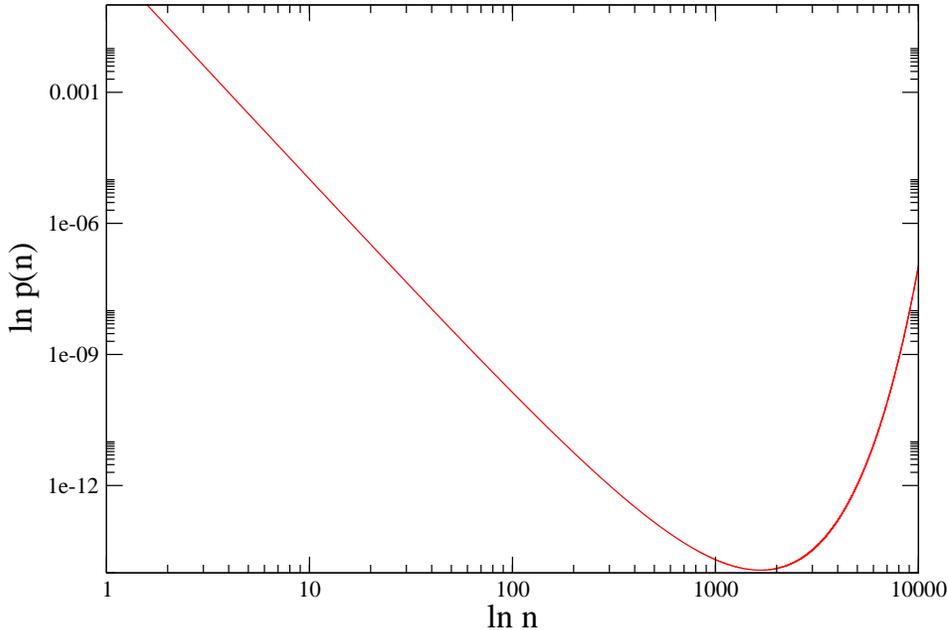}
\caption{ln-ln plot of the single-site distribution $p(n)$ given by
  (\ref{gc_p(n)}) v.\ particle
  number $n$. The curve shown is for $A/F(z)=1$, $\eta = 0.002$ and
  $b=5$, and should be compared with the diamonds in figure \ref{can_dist}.}
\label{gc_dist}
\end{center}
\end{figure}

However the condensate is not correctly reproduced by the introduction
of a cut-off. This is seen by noting that (\ref{p(n)gc}) becomes
\begin{equation} \label{gc_p(n)}
p(n) =\frac{A}{F(z)}\exp(n\eta - b\ln n)
\end{equation}
which has a minimum at $n=b/\eta$ and increases to a boundary maximum at the
maximum allowed value $n=N$, see figure \ref{gc_dist}.  
The piece of $p(n)$ to the right of the
minimum corresponds to the condensate.  However this piece is not
centred about the value $(\rho-\rho_c)L$ as the condensate should be.
This is because we have to take $z>1$ in order to account for the condensate
which implies $p(n)$ must increase at large $n$ rather than having a maximum
(as is seen in figure \ref{can_dist}).
This is in contrast to usual Bose condensation, in momentum space in the ideal
Bose gas, where the condensate is accounted for by taking $z \nearrow 1$
in the appropriate way \cite{Huang}.
To correctly describe the condensate in the present case one
has to work within the canonical ensemble \cite{EMZ05}.

\subsection{Condensation condition on hop rates $u(n)$}
\label{sec:u(n)cond}
So far our discussion of the condensation mechanism has centred on the
single-site weight function $f(n)$.  In most applications of ZRP,
however, one is given the hop rate $u(n)$ (see section
\ref{applications}), therefore we now consider the conditions for
condensation implied on $u(n)$.

For condensation we require that the infinite series $zF'(z) =
\sum_{n=1}^\infty nz^nf(n)$ converges for some value of $z$.  The
ratio of successive terms $n-1$, $n$ in the series is given by
\begin{equation}
\frac{(n-1) f(n-1)}{z\, n f(n)} =  \frac{u(n)}{z}\left(1 - \frac{1}{n}\right).
\end{equation}
The ratio test \cite{Arfken} tells us that for convergence of the
series this ratio should decay more slowly than $1 + 1/n$, thus for
condensation to occur, i.e. convergence at $z=\beta$, $u(n)$ should
decay more slowly than
\begin{equation}
u(n) \simeq \beta \left(1 + \frac{2}{n}\right) \quad\mbox{for}\quad n \gg 1\;.
\label{u(n)a}
\end{equation}
This corresponds to the asymptotic expression for $f(n)$ (\ref{f(m)a})
studied in the previous subsections.

If $u(n)$ decays, for example, as
\begin{equation}
u(n) \simeq \beta \left(1 +
\frac{a}{n^\lambda}\right)\quad\mbox{for}\quad n \gg 1\;, 
\end{equation}
where $0< \lambda <1$,  one obtains condensation but with the
fluid phase now distributed according to a stretched exponential
distribution $p(n) \sim \exp( -a n^{1-\lambda}/(1-\lambda))$.

If $u(n)$ increases as $n\to \infty$ the series has infinite radius of
convergence and so condensation never occurs.  On the other hand if
$u(n) \to 0$ as $n\to \infty$, the series will have zero radius of
convergence. This implies condensation at any density. Moreover the
density in the fluid phase will tend to zero, therefore the fraction
of particles in the condensate will tend to one.

These conditions on $u(n)$ lead directly to the criteria for phase
separation discussed in section \ref{sec:ps}. When the current of
particles out of a domain of size $n$ decays to an asymptotic value
$\beta$, for large $n$, more slowly than $\beta (1+2/n)$, phase
separation occurs above some critical particle density. This phase
separated state consists of a single domain containing a finite
fraction (less than one) of particles in the system.  On the other
hand, when the current of particles out of a domain of size $n$ decays
asymptotically to zero, then strong phase separation occurs leading to
a state in which a single domain contains a fraction of particles
equal to one --- the remainder of the system contains a finite number
of particles.

One can also infer the condensation phenomena that occur in the shaken
granular gases discussed in section \ref{sec:sgg}. In this case, both
the hop rates (\ref{eggers}) and (\ref{LD}) decay to zero as $n\to
\infty$. Thus, in the limit $N$, $L\to \infty$ with $\rho$ fixed, both
models will evolve to a steady state in which a single compartment
contains a fraction of particles
equal to one and the remainder of the system contains a finite number
of particles. The corresponding analysis in the limit $N\to\infty$
with $L$ fixed has been carried out in \cite{T04}.

\subsubsection{Explicitly solvable case}
\label{sec:sc}
It turns out that if one takes $u(n) = \beta(1+ b/n)$ $\forall n>0$
one can calculate the critical density, given by (\ref{rhoc}), exactly
\cite{GSS03,G03}.  In this case one has
\begin{equation}
f(n) = \beta^{-n} \prod_{i=1}^n \frac{i}{i+b} = \beta^{-n} \frac{n!}{(b+1)_n}
\end{equation}
where 
\begin{equation}
(a)_n = a(a+1)\ldots(a+n-1)\;,
\label{poch}
\end{equation}
is the Pochhammer symbol.
Thus
\begin{eqnarray}
F(z) &=& \sum_{n=0}^\infty (z/\beta)^n \frac{n!}{(b+1)_n}\
 = \sum_{n=0}^\infty (z/\beta)^n \frac{(1)_n (1)_n}{(b+1)_n\, n!}\nonumber\\
F'(z)&=& 
\frac{1}{\beta} \sum_{n=0}^\infty (z/\beta)^n \frac{(n+1)(n+1)!}{(b+1)_{n+1}}
 =
\frac{1}{\beta(b+1)} \sum_{n=0}^\infty (z/\beta)^n \frac{(2)_n
 (2)_n}{(b+2)_n\, n!}\nonumber 
\end{eqnarray}
and one finds using the hypergeometric function identity \cite{AAR}
\begin{equation}
\sum_{n=0}^\infty
\frac{ (a)_n\, (b)_n}{(c)_n n!} =
\frac{\Gamma(c)\Gamma(c-a-b)}{\Gamma(c-a)\Gamma(c-b)} \quad \mbox{for}
\quad c >a+b 
\end{equation}
that
  $F(\beta)= b/(b-1)$ and $F'(\beta)= b/(\beta(b-1)(b-2))$. Thus 
(\ref{rhoc}) yields
\begin{equation}
\rho_c = \frac{1}{b-2}\;.
\end{equation}
Note however that one does not expect $\rho_c$ to be universal i.e. it
generally depends on the form of $f(n)$ for small values of $n$, not
just on the asymptotic behaviour.

\subsection{Dynamics of condensation}
\label{sec:homdyn}
In this section we turn to the dynamics of the condensation
process. Starting from a homogeneous distribution of particles, the
condensate emerges as the result of a non-trivial coarsening process
\cite{OEC98}.  The excess particles accumulate on a number of sites
and the coarsening is then determined by the exchange of particles between
these sites; this results in their elimination and hence growth in the
mean particle number at such sites. The object of this section is
to determine the scaling satisfied by this growth.

We consider the case where the hops rates have the asymptotic
form
\begin{equation} \label{1+b/n}
u(n) = 1+b/n\;,
\end{equation}
when the system evolves from an initially random distribution of
particles at a supercritical density $\rho>\rho_c$. Two complementary
approaches have been taken to analyse the dynamics for this case; one
is a mean-field approach, and the other is a heuristic scaling
approach based on random walk arguments.

\subsubsection{Mean-field dynamics} 

The mean-field approach is due to Godr\`eche \cite{G03} and is based
on earlier studies of the dynamics of urn models \cite{DGC98, GL01}.
Here we outline this approach, and refer the reader to \cite{G03} for
details.

In the mean-field system we assume the steady state factorises,
and allow  particles to hop from any site to any
other in the system. Thus the probability $p_t(n)$ that a site
contains $n$ particles at time $t$ satisfies a Master equation given
by
\begin{equation} \label{mfme}
\frac{{\rm d} p_t(n)}{{\rm d}t} = u(n\!+\!1) p_t(n\!+\!1) + \langle
u_t\rangle p_t(n\!-\!1) - u(n) p_t(n) - \langle u_t \rangle p_t(n)\;,
\end{equation}
for $n>0$, where $\langle u_t \rangle = \sum_{m=1}^{\infty} u(m)
p_t(m)$ is the expectation value of the hop rate at time $t$, and
\begin{equation} \label{mfbc}
\frac{{\rm d} p_t(0)}{{\rm d}t} = u(1) p_t(1) - \langle u_t\rangle p_t(0)\;.
\end{equation}
These equations are non-linear, due to the dependence of $u_t$ of
$p_t(n)$, however the work of \cite{G03, DGC98, GL01} has
shown how one can obtain the late-time dynamics in the condensed phase
(and also at criticality, although we do not address this point
further here). One finds that the system decomposes into two distinct
parts: one which includes sites that have effectively reached a steady
state at density $\rho_c$, and the other which includes sites that
contain the  excess $(\rho-\rho_c)L$ particles that form the
condensate.

In the condensed phase, in the limit $t \to \infty$, $\langle
u_t\rangle \to \langle u\rangle$, the steady state expectation value
of the hop rate. In the steady state, $\langle u\rangle$ is equal to
the fugacity $z$, and in the condensed phase $z$ is given by its
maximum allowed value, $\beta$, determined by the radius of
convergence of the sum (\ref{Fdef}). For the rate (\ref{1+b/n})
$\beta=1$ therefore $\langle u\rangle=1$. Therefore for late times one
sets
\begin{equation} \label{ut}
\langle u_t \rangle \simeq 1 + A\epsilon_t\;,
\end{equation}
where $A$ is an amplitude and $\epsilon_t$ is a small time scale, both
to be determined later. Next, one observes that the terms in
(\ref{mfme}) which reduce the number of particles at a site are those
which depend on $u(n)$, and those which increase the number of
particles at a site depend on $\langle u_t\rangle$. 
Sites at which $n$ is finite reach an effective
steady state and the dynamics of the condensation is determined by
those sites at which $n$ is large (we will refer to sites at which $n$
is extensive as condensate sites). For sufficiently late times these
sites are in the scaling regime. Thus one defines the scaling variable
$x=n\epsilon_t$, and the probabilities $p_t(n)$ are rewritten in terms
of a scaling function $g(x)$ i.e.
\begin{equation} \label{simsol}
p_t(n) \simeq \epsilon_t^2 g(x)\;.
\end{equation}
Now one substitutes the three equations (\ref{1+b/n}), (\ref{ut}) and
(\ref{simsol}) into the Master equation (\ref{mfme}), which yields the
differential equation
\begin{equation} \label{g(x)}
g''(x)+\left( \frac{x}{2} - A +\frac{b}{x} \right) g'(x) +
\left(1-\frac{b}{x^2} \right) g(x)=0\;,
\end{equation}
satisfied by $g(x)$. As a byproduct one also obtains $\dot{\epsilon_t}
\sim \epsilon_t^3$, hence
\begin{equation}
\epsilon_t \sim t^{-1/2}\;.
\end{equation}
The amplitude $A$ is determined by the properties of $g(x)$ for small
and large $x$ and is found to be universal in the sense that it only
depends on $b$, therefore the scaling function $g(x)$ is also
universal in this sense. Properties of the solution of (\ref{g(x)})
for $g(x)$ can be found in \cite{G03,DGC98}.
  
The picture that emerges from this mean-field analysis is that the
typical number of particles at condensate sites scales as
$t^{1/2}$. Since the total number of particles at such sites is
$(\rho-\rho_c) L$, the number of condensate sites must scale as
$(\rho-\rho_c)L^{-1/2}$. Hence the mean condensate size $\langle m(t)
\rangle$, defined as the total number of particles at condensate sites
divided by the number of condensate sites at time $t$, obeys a scaling
law
\begin{equation} \label{scaling law}
\langle m(t) \rangle \sim t^{\delta}\;,
\end{equation}
defining the exponent $\delta$, where
\begin{equation}
\delta = 1/2\;.
\end{equation}
This result is exact for the ZRP in an infinite number of dimensions,
and is also expected to hold above an upper critical
dimension. Significantly, numerics indicate this is the correct result
in one dimension for asymmetric hopping dynamics. However, as
discussed in \cite{G03}, below the upper critical dimension both
dimensionality and bias in the hopping dynamics may be expected to
play an important role, and numerics indicate that for symmetric
dynamics the exponent changes to $\delta=1/3$. In order to address
these issues further we turn to an alternative, heuristic scaling
approach.

\subsubsection{Random walk argument}

This approach, given in \cite{GSS03}, is based on the
observation (also observed in the mean-field approach) that the
dynamics can be divided into distinct regimes: (i) a nucleation
regime, during which the excess $(\rho-\rho_c)L$ particles accumulate at a
finite number of condensate sites such that each condensate site
contains a number of particles of order $L$ --- the
remaining sites, which we refer to as bulk sites, have converged to
the steady state with density $\rho_c$; (ii) a coarsening regime,
during which the condensate sites exchange particles through the bulk
--- the bulk is viewed as a homogeneous background during this
process. The exchange of particles between condensate sites leads to
the growth of larger condensates at the expense of smaller ones. This,
in turn, leads to a decrease in the number of condensate sites and an
increase in the mean condensate size $\langle m(t)\rangle$. Thus the
coarsening regime persists until only a single condensate site
remains. The random walk argument yields a prediction of the exponent
$\delta$ appearing in the scaling law (\ref{scaling law}) which
determines the growth of $\langle m(t)\rangle$, the mean condensate
size.
 
We now outline how the random walk argument is applied to the ZRP in
one dimension to obtain predictions for the exponent $\delta$ for both
asymmetric dynamics and symmetric dynamics. During the coarsening
regime, the mean rate at which particles hop from bulk sites is given
by $\langle u \rangle = 1$. However at condensate sites particles
escape with a rate $u(n)-\langle u \rangle=b/n$, where $n =
\mathcal{O}[(\rho-\rho_c)L]$. Therefore a particle escapes from a
condensate site in a time that scales as
$\mathcal{O}(L)$. We must now find the time it takes
such a particle to reach the next condensate site. This time depends
on the symmetry of the hopping dynamics.
\begin{description}
\item[a) Asymmetric dynamics:] In this case, the particle reaches the
next condensate site to the right with certainty. 
The particle mobility is given by
the mean hop rate from occupied sites in the bulk, which is a finite
constant given by $\langle u \rangle/(1-p(0))$. The average distance
between two condensate sites scales as $\mathcal{O}(L)$ therefore the
typical time a particle spends in the bulk, having escaped from a
condensate site, scales as $\mathcal{O}(L)$. This time is of the same
order in $L$ as the typical escape time from condensate sites
therefore there are only a finite number of excess particles in the
bulk at any time. Thus the transit time between two condensate sites
does not limit the dynamics of the coarsening in any way. The growth
of the mean condensate size is determined only by the time it takes a
condensate site to lose all of its particles. This time scales as
$\mathcal{O}(L\times L)$ therefore the normalised mean
condensate size grows like
\begin{equation}
\frac{\langle m(t)\rangle}{(\rho-\rho_c)L} \sim \left( \frac{t}{\tau}
\right)^{1/2} \;,
\end{equation}
where the time scale for the coarsening regime, $\tau \sim
\mathcal{O}(L^2)$. Therefore the prediction $\delta=1/2$ for
asymmetric dynamics is the same as for the mean-field analysis.
 
\item[b) Symmetric dynamics:] In this case, having left a condensate
site, a particle performs a symmetric random walk with a diffusion
constant given by $\langle u \rangle/(1-p(0))$. Now, the probability
that a particle reaches the next condensate site, which is determined
by the solution of the `gamblers' ruin' problem \cite{feller}, is
proportional to the inverse separation of condensate sites. Thus this
probability is of order $L$, so particles almost
certainly return to the condensate site they have just left. The
typical time it takes a particle to escape from a condensate site is
therefore $\mathcal{O}(L^2)$. The transit time for such a particle is
given by the first passage time for a symmetric random walk which is
proportional to the square of the distance
i.e. $\mathcal{O}(L^2)$. Therefore, as in the asymmetric case, this
time is of the same order in $L$ as the typical escape time from
condensate sites so the growth of $\langle m(t)\rangle$ is determined
only by the time it takes a condensate site to lose all of its
particles. This time scales as $\mathcal{O}[L\times L^2]$ therefore
the normalised mean condensate size grows like
\begin{equation}
\frac{\langle m(t)\rangle}{(\rho-\rho_c)L} \sim \left( \frac{t}{\tau}
\right)^{1/3} \;,
\end{equation}
where the coarsening time scale $\tau \sim \mathcal{O}(L^3)$. Thus
$\delta = 1/3$ for symmetric dynamics.
\end{description}
The results of simulations are consistent with the predictions of the
random walk argument in both asymmetric and symmetric cases \cite{G03,
GSS03}.

The random walk argument can also be applied to the ZRP in more than
one dimension \cite{GH05}. The condensate sites may be viewed as
`trapping' sites amongst which the excess particles are exchanged. The
key feature of random walks in higher dimensions is that the
probability that a particle does not return to the condensate site it
has just vacated is $\mathcal{O}(1)$ for all walks except for the
symmetric random walk in two dimensions. For symmetric random walks in
two dimensions there are logarithmic corrections: the escape
probability is $\mathcal{O}[({\rm ln}L)^{-1}]$, where $L$ is now the
number of sites in the system. Therefore the upper critical dimension
for the coarsening of the ZRP is two and above two dimensions, we
expect the coarsening to be determined by the mean-field results for
the scaling function and exponent, $\delta=1/2$.

The coarsening time scales can be summarised as
follows
\begin{displaymath}
\tau \sim \left\{ 
\begin{array}{ll}
L^3 & {\rm in} \quad d=1\;,\\
L^{2} {\rm ln} L  & {\rm in} \quad d=2\;,\\
L^{2} & {\rm in} \quad d>2\;,
\end{array} \right.
\end{displaymath}
for symmetric dynamics, and
\begin{equation}
\tau \sim L^{2} \qquad \forall d\;,
\end{equation}
for asymmetric dynamics.

Note that mean-field analysis reproduces the correct exponent for the
asymmetric one-dimensional case.  The reason for this is that random
walk arguments yield the same result, $\delta=1/2$, whenever the
escape probability is of order $\mathcal{O}(1)$, and this is the
situation in the one dimensional case with asymmetric dynamics.


\section{Heterogeneous system}
\label{sec:hetero}
\subsection{Condensation}
We now to turn to mechanisms of condensation and the associated
coarsening dynamics in the heterogeneous ZRP.  In a heterogeneous
system the single-site weights $f_l(n)$ depend on the site $l$.
Taking the most general case discussed in section \ref{sec:arblat}
(where the hop rate from $l$ to $k$ is $u_l(n_l) W_{kl}$) we have
\begin{equation}
f_l(n) = s_l^n \prod_{i=1}^n \frac{1}{u_l(i)}\;.
\end{equation}
Working in the
grand canonical ensemble as in (\ref{sec:gce}) one finds
\begin{equation}
{\cal Z}_L(z) =  
\sum_{\{ m_l =0\}}^{\infty} z^{\sum_l m_l}
 \prod_{l=1}^{L} f_l (m_l) 
= \prod_l  F_l(z)
\label{Zgch}
\end{equation}
where 
\begin{equation}
F_l(z) = 
 \sum_{m=0}^{\infty} z^{m} \ f_l(m)\;, 
\label{hetFdef}
\end{equation}
and (\ref{rhogc}) becomes
\begin{equation}
\rho = \frac{z}{L} \sum_l \frac{F_l'(z)}{F_l(z)}\;.
\label{rhocongc}
\end{equation}

To simplify things we consider a one-dimensional lattice so that $s_l =1$
and we also take the hop rate of the form $u_l(n)=u_l$ i.e. there is no
dependence on the number of particles at the departure site.  In this
case $f_l$ is given by
\begin{equation}
f_l(n) = \left( \frac{1}{u_l}\right)^{\! n}\;.
\end{equation}
At this point the mapping to an ideal Bose gas is evident: the $N$
particles of the ZRP are viewed as Bosons which may reside in $L$
states with energies $E_{l}$ determined by the site hopping rates:
$\exp(-\beta E_{l}) = 1/u_{l}$.  Thus the ground state corresponds to
the site with the lowest hopping rate $u_{\rm min}$.

We can sum the geometric series (\ref{hetFdef}) to obtain
$F_l=1/(1-z/u_l)$ and
\begin{equation}
\rho = \frac{z}{L} \sum_l \frac{1}{u_l-z}
\label{rhocongc2}
\end{equation}
Thus the maximum allowed value of $z$ is $u_{\rm min}$.  In the large
$L$ limit we may proceed as in the usual theory of Bose condensation
\cite{Huang} and write
\begin{equation}
\rho = \frac{\langle n_{\rm min} \rangle}{L}
+  \int_{u_{\rm min}}^{\infty} du  {\cal P}(u) \  \frac{z}{u-z}
\label{gce2}
\end{equation}
where $\langle n_{\rm min} \rangle$ is the average number of particles
at the slowest site and ${\cal P}(u)$ is the probability distribution
of site hopping rates.  Interpreting ${\cal P}(u)$ as a density of
states, equation (\ref{gce2}) corresponds to the condition that in the
grand canonical ensemble of an ideal Bose gas the number of Bosons per
state is $\rho$.  If the density of states vanishes as $u\searrow
u_{\rm min}$ as
\begin{equation}
{\cal P}(u) \sim (u-u_{\rm min})^\gamma\quad \gamma >0
\end{equation}
then the second term in (\ref{gce2}) will be finite as $z \nearrow
u_{\rm min}$ and will define $\rho_c$.  For $\rho >\rho_c$, the first
term in (\ref{gce2}) must be finite thus $\langle n_{\rm min} \rangle
\sim O(L)$.
This is the mechanism of condensation underlying heterogeneous models
of networks, for example \cite{BB01}.

\subsection{Single defect Site}
\label{sec:sds}
A very simple example, which
serves to illustrate the mechanism of condensation,
 is to have just one `slow site' i.e. $u_1=p <1$
while the other $L-1$ sites have hopping rates $u_{l}=1$ when $l>1$
\cite{E96}.
The simplicity of this system actually allows the canonical partition
function (\ref{Z_L,N}) to be calculated. This is easy to write down as
\begin{equation}
 Z_{L,N}  =  \sum_{n=0}^N  p^{-n} \sum_{n_2 \ldots n_L}
\delta(  \sum_{\mu=2}^L  n_\mu{-}N+n)
\end{equation}
where $n$ counts the number of particles on the slow site.
Identifying the sum over $n_2 \ldots n_L$ as the number of ways of
distributing $N-n$ indistinguishable balls in $L-1$ boxes implies
\begin{equation}
 Z_{L,N}  =  \sum_{n=0}^{N}
{N{+}L{-}n{-}2 \choose L{-}2}
p^{-n} \; .
\label{z1slow}
\end{equation}

In order to determine which terms dominate the sum (\ref{z1slow}) we
look for the stationary point of the summand, which occurs when the
ratio of consecutive terms $n,n-1$ is equal to one:
\begin{equation}
\frac{1}{p}\frac{N-n+1}{(N+L-n-1)}=1\;.
\end{equation}
For $L$ large, we find a solution $n\simeq L(\rho-\rho_c)$ when $\rho
> \rho_c= p/(1-p)$. For $\rho < \rho_c$, however, we do not find a
solution for positive $n$ therefore the maximum is the boundary term
$n=0$ and the sum will be dominated by terms $n \sim {\mathcal O}(1)$.
Thus $\rho_c=p/(1-p)$ is the critical density above which we have a
condensate: we see clearly that in the condensed phase the defect site
serves to absorb the excess number of particles $L(\rho-\rho_c)$; in
the fluid phase the number of particles at the defect site is
${\mathcal O}(1)$.

In the fluid phase we evaluate the asymptotic behaviour of the
partition function by expanding for small $n$, since these terms
dominate the sum:
\begin{equation}
 Z_{L,N}  \simeq  \sum_{n=0}^{N}
{N{+}L \choose L}
\frac{L^2 N^n}{(N+L)^{n+2}} p^{-n} 
\simeq
{N{+}L \choose L}
\frac{\rho_c}{(1+\rho)(\rho_c-\rho)}
\; .
\label{z1f}
\end{equation}
In the condensed phase one approximates the sum by an integral and uses
Stirling's formula to obtain
\begin{eqnarray}
\lefteqn{ Z_{L,N}  \simeq   L^{1/2}
\int_0^\rho \frac{dx}{\sqrt{2\pi}}
\left(\frac{1+\rho-x}{\rho-x}\right)^{1/2}}
\\\hspace{2cm}
 \times\exp  L\left[ (1+\rho-x)\ln(1+\rho-x) - (\rho-x)\ln(\rho-x)
   -x\ln p\right]\nonumber 
\end{eqnarray}
where $x= n/L$.  Then the integral is dominated by $x \simeq
\rho-\rho_c$ and expanding to second order and performing the Gaussian
integral yields
\begin{equation}
 Z_{L,N}  \simeq   (1-p)^{-(L+1)}p^{-N}\;. \label{z2f}
\end{equation}
From expressions (\ref{z1f}, \ref{z2f}) and (\ref{speed}), the mean
hop rate in the two phases is given by
\begin{eqnarray}
\langle u \rangle = \frac{\rho}{1+\rho} \quad \rho < p/(1-p)\;,\\
\langle u \rangle = p \qquad \mbox{for} \quad \rho > p/(1-p)\;.
\end{eqnarray}
These expressions for $\langle u\rangle$ are easily understood as the
hop rate being limited by the condensate in the condensed phase and by
the probability of empty sites, $p(0)= 1/(1+\rho)$, in the fluid
phase.

\subsubsection{Defect with $n$-dependent hop rate}
Recently the case of a single defect site with an occupation number
dependent hop rate $u_1(n)$ has been considered.  This allows one to
investigate the interplay between condensation driven by particle
number dependent hop rates and condensation driven by heterogeneity
\cite{AEM04}.

Again one can write down the canonical partition function.  Taking
$$
u_1(n) = p\left(1 + \frac{b}{n}\right) \quad \mbox{for}\quad n>0
$$
with $p<1$ and $u_l(n)=1$ for $l\neq 1$ yields
\begin{equation}
Z_{L,N} = \sum_{n=0}^N p^{-n} \frac{n!}{(1+b)_n}
{N+L-n-2 \choose L-2}\;,
\end{equation}
where $(1+b)_n$ is a Pochhammer symbol defined in (\ref{poch}).
One can again determine the stationary point of the summand by setting
the ratio of consecutive terms equal to one, resulting in a quadratic
expression
\begin{equation}
\frac{n}{p(b+n)} \frac{(N-n+1)}{(N+L-n-1)}=1
\end{equation}
A real, nonnegative solution of this quadratic exists for
\begin{equation}
\rho> \rho_2 = \frac{p}{1-p} + \frac{2}{1-p} \left( \frac{bp}{L}\right)^{1/2}
+ {\mathcal O} (1/L)
\end{equation}
As $L \to \infty$ one has a condensation transition at $\rho_c =
p/(1-p)$, which is independent of $b$. Thus the condensation is
triggered by the site being slow rather than the $n$ dependence of
$u_1(n)$.  However on any large but finite system (with $b>0$) the
condensed phase does not appear until $\rho_2 = \rho_c + {\mathcal
O}(L^{-1/2})$. Thus on a finite system the fluid phase continues above
the thermodynamic critical density.  This also results in the mean
hop rate going above its predicted thermodynamic value.  At
$\rho=\rho_2$ one sees the system wander between a fluid state and a
condensed state \cite{AEM04}.

\subsection{Dynamics}
\label{sec:het dyn}
The coarsening behaviour of the heterogeneous ZRP can be obtained by
using a scaling argument to predict the growth law for the typical
condensate size.  We consider the case where the hop rates are drawn
from the distribution with support $u \in [ u_{\rm min}, 1]$
\begin{equation}
{\cal P}(u) = \frac{1+\gamma}{(1-u_{\rm min})^{\gamma+1}} (u-u_{\rm
min})^\gamma\;,\qquad \gamma >0\;.
\end{equation}
The argument is rather similar to the homogeneous case. After an
initial nucleation regime, most of the system relaxes to the steady
state except for a finite number of sites which contain the excess
particles in the system. Eventually, only two condensate sites will
remain and these will be the sites with the two slowest hop
rates. These two sites will be separated by a distance of order
$\mathcal{O}(L)$ (if they were separated by a distance of order
$\mathcal{O}(1)$ they would effectively act as a single slow site).
This implies that the second slowest site must contain a number of
particles of order $\mathcal{O}(L)$, since it is the slowest site
encountered by particles in a finite fraction of the system. This is
to be compared with the steady state, in which the $j$-th slowest site
contains a number of particles of order ${\cal
O}((L/j)^{1/(1+\gamma)})$ which is subextensive for $\gamma >0$.  This
scaling can be inferred from an extremal statistics argument.
 
In the case of asymmetric dynamics \cite{KF96,K00,JB03}, the second
slowest site gains particles from the left with rate $u_{\rm min}$,
but loses particles to the right with rate $u_1<u_{\rm min}$, thus it
loses particles with a net rate $\Delta u = u_1-u_{\rm min}\sim {\cal
O}(L^{-1/(1+\gamma)})$. Therefore the time $\tau$ this site takes to lose
all its excess particles (i.e. excess above its typical steady state
occupancy) scales as ${\cal O}(L/\Delta u)$. Thus
\begin{equation}
\tau \sim L^{(2+\gamma)/(1+\gamma)}\;,
\end{equation}
and the normalised mean condensate size grows like
\begin{equation}
\frac{\langle m(t)\rangle}{(\rho-\rho_c)L} \sim \left( \frac{t}{\tau}
\right)^{\frac{1+\gamma}{2+\gamma}} \;,
\end{equation}
so the exponent $\delta = (1+\gamma)/(2+\gamma)$ for asymmetric
dynamics.

For symmetric dynamics a similar argument can be used except that now 
the coarsening timescale $\tau$ is augmented by a factor $L$ due to the
probability that a particle returns to the condensate site it has just
vacated rather that reaching the slowest site \cite{BJ02}. Thus
\begin{equation}
\tau \sim L^{(3+2\gamma)/(1+\gamma)}\;,
\end{equation}
and the normalised mean condensate size grows like
\begin{equation} \label{BJm(t)}
\frac{\langle m(t)\rangle}{(\rho-\rho_c)L} \sim \left( \frac{t}{\tau}
\right)^{\frac{1+\gamma}{3+2\gamma}} \;,
\end{equation}
so the exponent $\delta = (1+\gamma)/(3+2\gamma)$ for symmetric
dynamics. This was shown for $\gamma >1$ in \cite{BJ02}, however
it in fact  holds for $\gamma >0$ \cite{KJ}. 


\section{Generalisations}
\label{generalisations}

\subsection{Two-species ZRP}
\label{sec:2szrp}
In view of the recent interest in two component models \cite{S03} it
is natural to generalise the ZRP to a model with two species of
conserved particles. In this way, one can explore the role of
conservation laws in the ZRP. In particular, considering the generic
nature of the condensation mechanism in the single species ZRP, it is
of interest to look for other mechanisms of condensate formation which
may have some generic applicability. Moreover, as discussed in section
\ref{applications}, the behaviour of shaken granular gases in which
the grains come in one of two sizes (i.e. a bidisperse system), and the
behaviour of networks with directed edges, can be understood in terms
of a two species ZRP. The particle dynamics we consider are chosen
such that the evolution of the two particle species are coupled, and the
nature of the coupling can be chosen such that condensation of one of
the species is induced by the other. Such an interplay arises in
models of particles moving on an evolving disordered background
\cite{LR97,DK00,DB00}. We discuss how this interplay leads to new
mechanisms of condensation and rich coarsening behaviour.

In one dimension, we define the two-species ZRP on a lattice
containing $L$ sites and with periodic boundary conditions. At each
site $l$, there are $n_l$ particles of species $A$ and $m_l$ particles
of species $B$. The total number of species $A$ particles in the
system is $N$ and the total number of species $B$ particles is
$M$. These particles hop from site $l$ to the neighbouring site to the
right, species $A$ with rate $u(n_l,m_l)$ and species $B$ with rate
$v(n_l,m_l)$.

\subsubsection{Steady state factorisation condition}

Happily, one of the most useful features of the single-species model
--- the factorised steady state --- remains a feature of the
two-species model provided the dynamics satisfy a certain constraint
\cite{GS03, EH03}. That is, one can still express the probabilities
$P(\{n_l\};\{m_l\})$ in the factorised form:
\begin{equation} \label{P(n,m)}
P(\{n_l\};\{m_l\}) = Z_{L,N,M}^{-1} \prod_{l=1}^{L} f(n_l,m_l) \;,
\end{equation}
provided the hop rates satisfy the constraint 
\begin{equation} \label{CE}
\frac{u(n_l,m_l)}{u(n_l,m_l\!-\!1)} = \frac{v(n_l,m_l)}{v(n_l\!-\!1,m_l)} \;,
\end{equation}
for $n_l,m_l \neq 0$. The choices of $u(n_l,0)$ and $v(0,m_l)$ remain
unconstrained. When the hop rates satisfy (\ref{CE}), we can find
single-site weights $f(n,m)$ given by
\begin{equation} \label{f(n,m)}
f(n,m) = \prod_{i=1}^{n} \left[ u(i,m) \right]^{-1}
\prod_{j=1}^{m} \left[ v(0,j) \right]^{-1} \;,
\end{equation}
and with $f(0,0) = 1$. We remark that the hop rates do indeed play a
symmetric role in (\ref{f(n,m)}), but that this symmetry is obscured
within the constraint equation (\ref{CE}). Finally, the normalisation
$Z_{L,N,M}$, equivalent to the canonical partition function, is given
by
\begin{equation} \label{Z_L,N,M}
Z_{L,N,M} = \sum_{\left\{ n_l \right\} , \left\{ m_l \right\}}
\prod_{l=1}^{L} f(n_l,m_l) \delta \left( \sum_{l=1}^L n_l - N \right)
\delta\left( \sum_{l=1}^L m_l - M\right)\;, 
\end{equation}
where the $\delta$-functions ensure that we only sum over
configurations with $N$ particles of species $A$ and $M$ of species
$B$.
 
The steady state (\ref{P(n,m)}) to (\ref{Z_L,N,M}) for this model can
be derived in much the same way as in the single species case. One
begins by writing down the steady state condition on the probabilities
$P(\{n_l\};\{m_l\})$. This has the form
\begin{eqnarray} \label{2specME}
\fl 0 = \sum_{l=1}^L \left[ \left\{ u(n_{l-1}\!+\!1,m_{l-1})
P(\ldots,n_{l-1}\!+\!1,n_l\!-\!1,\ldots;\ldots,m_{l-1},m_l,\ldots)
\right. \right. \nonumber \\
\left. - u(n_l,m_l) P(\ldots,n_{l-1},n_l,\ldots;\ldots,m_{l-1},m_l,\ldots)
\right\} \theta(n_l) \nonumber \\ 
+ \left\{ v(n_{l-1},m_{l-1}\!+\!1)
P(\ldots,n_{l-1},n_l,\ldots;\ldots,m_{l-1}\!+\!1,m_l\!-\!1,\ldots)
\right. \nonumber \\ 
\left. \left. - v(n_l,m_l)
P(\ldots,n_{l-1},n_l,\ldots;\ldots,m_{l-1},m_l,\ldots) \right\}
\theta(m_l) \right]
\end{eqnarray}
where the Heaviside function $\theta(x_l)$ ensures that site $l$ is
occupied in order for a particle either to have arrived or to be able
to vacate there. Now, the first term on the rhs of (\ref{2specME}) is
a gain term due to an $A$ particle hopping into site $l$ from site
$l\!-\!1$ and the second term is a loss term due to an $A$ particle
hopping out of site $l$; the third and fourth terms represent
analogous processes for the $B$ particles. As in the single-species
case, we simply insert the factorised form (\ref{P(n,m)}) into
(\ref{2specME}), but now we ask that the gain and loss terms due to
the dynamics of the $A$ particles cancel independently of the gain and
loss terms due to the dynamics of the $B$ particles. We look to
achieve this cancellation for each term $l$ in the sum separately
hence 
\begin{equation}
\fl u(n_l,m_l) f(n_{l-1},m_{l-1})
f(n_l,m_l) = u(n_{l-1}\!+\!1,m_{l-1})f(n_{l-1}\!+\!1,m_{l-1})
f(n_l\!-\!1,m_l) \;,
\end{equation}
for all $n_l \neq 0$, and  
\begin{equation}
\fl v(n_l,m_l) f(n_{l-1},m_{l-1})
f(n_l,m_l) = v(n_{l-1},m_{l-1}\!+\!1)f(n_{l-1},m_{l-1}\!+\!1)
f(n_l,m_l\!-\!1) \;,
\end{equation} 
for all $m_l \neq 0$, for all values of $l$ and after cancelling
common factors. These equations in turn imply that 
\begin{eqnarray} \label{cancelAs}
\frac{u(n_l,m_l) f(n_l,m_l)}{f(n_l\!-\!1,m_l)} =
\frac{u(n_{l-1}\!+\!1,m_{l-1})f(n_{l-1}\!+\!1,m_{l-1})}{f(n_{l-1},m_{l-1})}
= {\rm constant} \;, \\
\label{cancelBs}
\frac{v(n_l,m_l) 
f(n_l,m_l)}{f(n_l,m_l\!-\!1)} =
\frac{v(n_{l-1},m_{l-1}\!+\!1)f(n_{l-1},m_{l-1}\!+\!1)}{f(n_{l-1},m_{l-1})}
= {\rm constant} \;,
\end{eqnarray}
for all $l$. Both constants are set equal to unity without loss of
generality, since they only appear as overall factors in the
normalisation. The two relations (\ref{cancelAs}) and (\ref{cancelBs})
are iterated to yield (\ref{f(n,m)}). However, (\ref{cancelAs}) and
(\ref{cancelBs}) also imply the constraint (\ref{CE}) on the choice of
$u(n_l,m_l)$ and $v(n_l,m_l)$.  To obtain this constraint, we use
(\ref{cancelAs}) and (\ref{cancelBs}) to obtain two expressions for
$f(n_l,m_l)$ in terms of $f(n_l\!-\!1,m_l\!-\!1)$:
\begin{equation} \label{constrained}
f(n_l,m_l) = \frac{f(n_l\!-\!1,m_l\!-\!1)}{u(n_l,m_l) v(n_l\!-\!1,m_l)}
= \frac{f(n_l\!-\!1,m_l\!-\!1)}{u(n_l,m_l\!-\!1) v(n_l,m_l)}\;.
\end{equation}
Since both of these expressions must give the same result the hop
rates are required to obey the constraint (\ref{CE}). Equation
(\ref{constrained}) also indicates how we might interpret physically
the constraint (\ref{CE}): it suggests that the relationship between
$f(n_l\!-\!1,m_l\!-\!1)$ and $f(n_l,m_l)$ is the same when an $A$
particle hops from $l$ and then a $B$, as when the $B$ particle hops
before the $A$. In other words the single-site weight $f(n_l,m_l)$ is
independent of the order in which the particle species arrived at $l$.

As in the single-species case, it is straightforward to generalise the
derivation of the steady state to an arbitrary lattice, including
generalisations to disorder in the hop rates, partial asymmetry in the
dynamics and any dimension. It is also possible to generalise to any
number of species, $Q$ say, in which case each species satisfies $Q-1$
constraints of the form (\ref{CE}), one for every pair of species
\cite{GS03}. 

\subsubsection{Condensation}

We now consider how the interaction of the two species may lead to a
condensation transition as a function of the particle densities,
\begin{equation}
\rho_A = \frac{N}{L} \qquad {\rm and} \qquad \rho_B = \frac{M}{L}\;,
\end{equation}
for species $A$ and $B$ respectively.  We consider a case where the
dynamics of one of the particle species --- the $B$ particles ---
depends only on the number of particles of the other species at the
departure site. Hence we take
\begin{equation} \label{v(n,m)}
v(n,m)=1+\frac{c}{(n+1)^\gamma}\;,
\end{equation}
for all $n$ and for $m>0$, where $c$, $\gamma>0$. This choice for
$v(n,m)$ now determines the $m$-dependence we must take for $u(n,m)$
through the constraint (\ref{CE}):
\begin{equation} \label{u(n,m)}
u(n,m) = \left( \frac{1+\frac{c}{(n+1)^\gamma}}{1+\frac{c}{n^\gamma}}
\right)^m\,.   
\end{equation}
We could choose to multiply $u(n,m)$ by some function of $n$ and still
satisfy (\ref{CE}) but we will proceed with just the rate
(\ref{u(n,m)}) for simplicity. The single-site weights $f(n,m)$,
determined by (\ref{f(n,m)}), are given by
\begin{equation} \label{f(n,m)a}
f(n,m) = \left( 1+\frac{c}{(n+1)^\gamma} \right)^{-m}\;.
\end{equation}

As in the single-species case, it is simplest to work within the grand
canonical ensemble in order to demonstrate condensation. The grand
canonical partition function is defined as
\begin{equation} \label{Z_L(z,y)}
\mathcal{Z}_L(z,y) = \sum_{n=0}^\infty \sum_{m=0}^\infty z^n y^m
Z_{L,n,m}\;,
\end{equation}
where we now have two fugacities, $z$ and $y$, which fix the
particle densities through 
\begin{equation} 
\rho_A = \frac{z}{L} \frac{\partial {\rm
    ln}\mathcal{Z}_L(z,y)}{\partial z} \;, \qquad 
\rho_B = \frac{y}{L} \frac{\partial {\rm ln}\mathcal{Z}_L(z,y)}{\partial y} \;.
\end{equation}
Now, we use (\ref{Z_L,N,M}) and perform the sums over $n$ and $m$ in
(\ref{Z_L(z,y)}) to obtain
\begin{equation}
\mathcal{Z}_L(z,y) = \sum_{\{ n_l \},\{ m_l\}} z^{\sum_l n_l} y^{\sum_l
  m_l} \prod_{l=1}^{L} f(n_l,m_l) = [F(z,y)]^L\;,
\end{equation}
where
\begin{equation} \label{F(z,y)}
F(z,y) = \sum_{n=0}^{\infty} \sum_{m=0}^\infty z^n y^m f(n,m)\;,
\end{equation}
and the densities are determined by
\begin{equation} \label{rhoA,rhoB}
\rho_A = z \frac{\partial {\rm ln} F(z,y)}{\partial z}  \;, \qquad
\rho_B = y \frac{\partial {\rm ln} F(z,y)}{\partial y}  \;.
\end{equation}
In the single species case, the critical density was determined by the
properties of $F(z)$ and its derivative when $z$ assumed its maximum
value $\beta$. A similar picture emerges here, which we illustrate for
$f(n,m)$ given by (\ref{f(n,m)a}).

\begin{figure} 
\begin{center}
\includegraphics[scale=0.5]{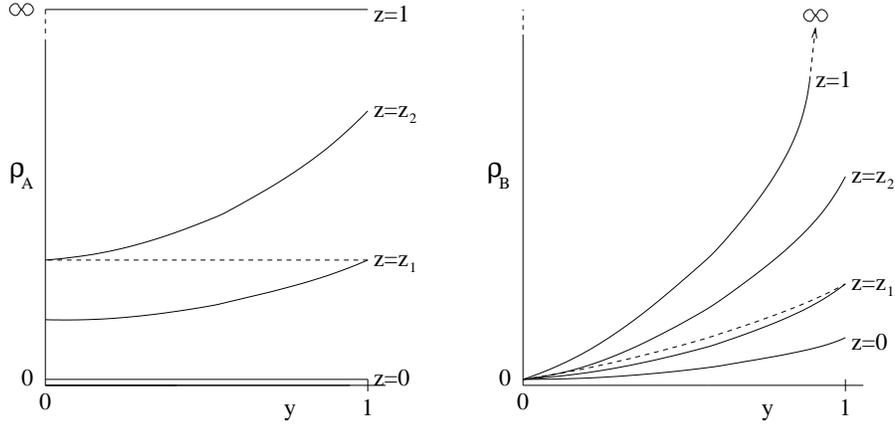}
\end{center} 
\caption{The solid lines represent schematic dependences of the
particle densities $\rho_A$ and $\rho_B$ for fixed values of $z$ and
as a function of $y$. The dashed line in the right hand graph
illustrates how $\rho_B$ varies as a function of $z$ and $y$ given
that $\rho_A$ is fixed.}
\label{fig:densities}
\end{figure}

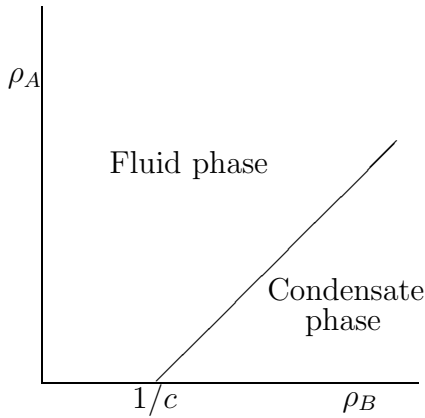
\begin{figure}
\begin{center}
\unitlength=1mm
\linethickness{0.4pt}
\begin{picture}(55,55)

\put(5,5){\line(1,0){50}}
\put(5,5){\line(0,1){50}}
\put(20.2,5.2){\line(1,1){32}}

\put(14,33){Fluid phase}
\put(35,16){Condensate}
\put(40,12){phase}

\put(0.5,45){$\rho_A$}
\put(45,2){$\rho_B$}

\put(17,1.6){$1/c$}

\end{picture}
\caption{Phase diagram for the two-species zero-range process with
$\gamma=1$.}
\label{fig:phase diagram} 
\end{center}
\end{figure}

Using (\ref{f(n,m)a}) in (\ref{F(z,y)}), and performing the sum over
$m$, one obtains 
\begin{eqnarray}
\label{fzy}
F(z,y) = \sum_{n=0}^{\infty} z^n
\frac{(1+n)^\gamma+c}{(1+n)^\gamma(1-y)+c}\,,\\ 
\label{zdzfzy}
z \frac{\partial}{\partial z} F(z,y) = \sum_{n=0}^{\infty} n z^n
\frac{(1+n)^\gamma+c}{(1+n)^\gamma(1-y)+c}\,, \\
\label{ydyfzy}
y \frac{\partial}{\partial y} F(z,y) = \sum_{n=0}^{\infty} (1+n)^\gamma z^n
\frac{y((1+n)^\gamma+c)}{[(1+n)^\gamma(1-y)+c]^2}\,.
\end{eqnarray}
The radii of convergence of these series are $z=1$ and $y=1$. We can
now use equations (\ref{fzy}) to (\ref{ydyfzy}) in (\ref{rhoA,rhoB})
to find $z$ and $y$, given values for $\rho_A$ and $\rho_B$. In figure
\ref{fig:densities}, we illustrate how $\rho_A$ and $\rho_B$ depend on
$y$ for fixed $z$.
This figure is obtained by noting that for a given $z$, both $\rho_A$
and $\rho_B$ are monotonically increasing functions of $y$. Then we
can infer the behaviour shown in figure \ref{fig:densities} by
analysing (\ref{rhoA,rhoB}) in the four limits $y\to 0$, $z\to 0$,
$y\to 1$, and $z\to1$. The existence of a condensation transition is
demonstrated by considering a system containing a density $\rho_A$ of
$A$ particles, represented by the dashed lines in figure
\ref{fig:densities}. The solution of (\ref{rhoA,rhoB}) requires that
$z$ and $y$ lie in the range $z_1 \leq z \leq z_2$ and $0 \leq y \leq
1$. In this range, $\rho_B$ increases monotonically from $\rho_B=0$,
where $z=z_2$ and $y=0$, to its maximum value, where $y=1$ and
$z=z_1$.  Thus, for every finite value of $\rho_A$ there exists a
finite maximum value of $\rho_B$ above which we can no longer solve
(\ref{rhoA,rhoB}).  When $\rho_B$ exceeds this maximum
(\ref{rhoA,rhoB}) cannot be satisfied and a condensation transition
ensues. Hence we obtain the phase diagram shown in figure
\ref{fig:phase diagram}.
For $\gamma=1$ the critical line is given simply by
\begin{equation} \label{critical line}
\rho_B = (1+\rho_A)/c \,.
\end{equation}

Thus the model exhibits a phase transition between a fluid phase, in
which both particle species are distributed exponentially, and a
condensed phase, in which $\rho_B$ exceeds a maximum critical value
and the excess $B$ particles condense onto a single site.  The number
of $B$ particles in this condensate is proportional to the system size
$L$ --- in the limit $N$, $M$, $L\to \infty$ it contains a finite
fraction of an infinite number of particles. However, we note that for
$m \to \infty$, $u(n,m) \to 0$ if $n$ is finite. But this is not
consistent with continuity since the current of $A$ particles away
from the condensate site is finite therefore $u(n,m)$ at the
condensate site must also be finite. Thus it must be that $n \to
\infty$ at the condensate in order that the current out of the
condensate site is finite. For large $n$ and $m$, $u(n,m) \sim {\rm
exp}(-\gamma m/n^{1+\gamma})$ and since $m$ is a number of order $m \sim
\mathcal{O}(L)$ at the condensate site, $n$ must be of order
$n \sim \mathcal{O}(L^{1/(1+\gamma)})$ for finite $u(n,m)$. Thus the
condensate of $B$ particles is sustained by a subextensive condensate
of $A$ particles.

The conditions on $u(n,m)$ and $v(n,m)$, and therefore $f(n,m)$ also,
leading to condensation have been considered for general rates in
\cite{HE04}.  Additional condensation transitions can be obtained by
supplementing the $A$ particle hop rate (\ref{u(n,m)}) with a factor
$(1+b/n)$. In this case two additional condensed phases arise, one in
which only the $A$ particles condense, and one in which both particle
species condense --- thus this phase also exhibits induced
condensation of one species by the other.

\subsubsection{Dynamics of condensation}
\label{sec:2sdyn}
The coarsening dynamics in the condensed phase shown in figure
\ref{fig:phase diagram} can be obtained by adapting the random walk
arguments given for the single species case \cite{GH}.

Thus, as before, we consider a system most of which has relaxed to the
steady state, except for a finite number of condensate sites. The
condensate sites contain a number of $A$ particles of order $n \sim
\mathcal{O}(L^{1/(1+\gamma)})$ and a number of $B$ particles of order
$m \sim \mathcal{O}(L)$, i.e. $m \sim n^{1+\gamma}$. Therefore, from
(\ref{v(n,m)}), the $B$ particle hop rate decays to its steady state
value as $c/m^{\gamma /(1+\gamma)}$, so the typical timescale over
which $B$ particles escape from condensate sites scales as $L^{\gamma/
(1+\gamma)}$. Then the time it takes a condensate site to lose all of
its $B$ particles scales as $L^{1+\gamma/ (1+\gamma)}$. This sets the
coarsening timescale, thus the growth of the mean $B$ particle
condensate size follows the scaling law (\ref{scaling law}) with
exponent
\begin{equation}
\delta = \frac{1+\gamma}{1+2\gamma}\;.
\end{equation}

A prediction for the exponent $\delta$ can also be obtained in the
case that the dynamics are symmetric. The dynamics are altered in the
same way as in the single species case, in that only every $L$-th
particle on average reaches the next condensate site rather than
returning to the one it has just departed, thus the typical
timescale over which $B$ particles escape from condensate sites is a
factor of $L$ longer: it scales as $L^{1+\gamma/(1+\gamma)}$.
Therefore the time it takes a condensate site to lose all of its $B$
particles scales as $L^{2+\gamma/ (1+\gamma)}$ and the coarsening
exponent is given by
\begin{equation}
\delta = \frac{1+\gamma}{2+3\gamma}\;.
\end{equation}

It is interesting to note that these exponents are the same as those
obtained for the heterogeneous ZRP with hop rates drawn from the
distribution (c.f.\ section \ref{sec:het dyn})
\begin{equation} \label{P(u)}
\mathcal{P}(u) =
[(\gamma^{-1}+1)/(1-u_{\rm min})^{\gamma^{-1}+1}](u-u_{\rm
  min})^{\gamma^{-1}}\;.     
\end{equation}
To consider why this might be the case, one views the dynamics defined
in (\ref{v(n,m)},\ref{u(n,m)}) as a model of particles (the $B$
particles) moving on an evolving disordered background (given by the
$A$ particles). By the time the coarsening regime has been reached, at
the condensate sites the evolving disorder is effectively
quenched. Therefore it is not necessarily surprising that the two
models exhibit similar coarsening behaviour for some distribution
$\mathcal{P}(u)$. The reason the form (\ref{P(u)}) is the relevant one
for the rates (\ref{v(n,m)},\ref{u(n,m)}) is as follows. In the
heterogeneous system, as discussed in section \ref{sec:het dyn}, the
coarsening is governed by the difference between the hop rates at the
two slowest sites in the system, $\Delta u$.  For the distribution
(\ref{P(u)}), $\Delta u \sim L^{-1/(1+\gamma^{-1})}$. This rate
separation plays the role of the background of $A$ particles in the
two-species model. The remaining contributions to the coarsening time
scale in the heterogeneous model are then determined by the symmetry
of the hopping dynamics, i.e.\ the coarsening time scale is given by a
factor of order $L$ for asymmetric dynamics, or a factor of order
$L^2$ for symmetric dynamics, multiplied by the inverse rate
separation. This leads to the same exponents as those obtained in the
condensed phase of the two-species model.

\subsection{Urn models and the Misanthrope process}
\label{sec:Mis}
A natural extension of the ZRP dynamics is to hop rates which depend
not only on the occupancy of the departure site, but also on that of
the target site.  For example, a recent application of such models is
to rewiring dynamics in networks as we discuss in section
\ref{sec:rewire}.  Models of this kind are often referred to as `urn
models'.  As reviewed in \cite{GL02} urn models comprise balls
distributed amongst a number of boxes with conserving dynamics for the
exchange of balls.  These dynamics are usually defined on a
fully-connected geometry and obey detailed balance with respect to
some energy function defined as a sum of single-site energies.

On the other hand the Misanthrope process, introduced in the
mathematical literature in \cite{C85}, is a model with dynamics
defined without reference to an energy function. In one dimension its
dynamics are defined in exactly the same way as the ZRP except that
now the hop rate depends on the occupancy of both departure and target
sites.  Thus a particle hops from site $l$ to site $l\! +\! 1$ with a
rate $u(n_l,n_{l\! +\! 1})$.

\subsubsection{Steady state factorisation condition}
It turns out that the steady state still has a
factorised form, provided the hop rates satisfy a constraint.
To see this, consider the condition on the steady state probabilities
$P(\{n_l\})$: 
\begin{eqnarray} \label{MEmis}
0 = \sum_{l=1}^{L} & \left[ u(n_{l-1}\!+\!1, n_l\!-\!1 )
  P(\cdots,n_{l-1}\!+\!1, n_l\!-\!1, \cdots) \theta(n_l) \right. \nonumber \\
  & \left. -
  u(n_{l-1},n_l) P(\{n_l\}) \theta(n_{l-1}) \right] \;.
\end{eqnarray}
To look for a factorised steady state of the form (\ref{P(C)}) we
introduce a counterterm which cancels under the sum and then look to
cancel each term in the sum separately, hence
\begin{eqnarray} \label{cancellation}
\fl \bar{f}(n_{l-1}) f(n_l) - f(n_{l-1})\bar{f}(n_l) &=&
  u(n_{l-1}\!+\!1, n_l\!-\!1 ) f(n_{l-1}\!+\!1) f(n_l\!-\!1)
  \theta(n_l) \nonumber \\ && -
  u(n_{l-1},n_l) f(n_{l-1}) f(n_l) \theta(n_{l-1})\;,
\end{eqnarray}
having cancelled common factors (a product over the functions $f(n_k)$
at all sites $k\neq l\!-\!1$, $l$), where $\bar{f}(n)$ is some
function to be determined. The lhs of (\ref{cancellation}) represents
the counterterm. We remark that this cancellation mechanism is
identical to that of the so-called `Matrix Product Ansatz' \cite{DEHP93}
in the case where matrices are replaced by the functions $f(n)$ and
$\bar{f}(n)$ \cite{ZS}.

In order to satisfy (\ref{cancellation}) for all values of $n_{l-1}$
and $n_l$ there are now three cases to consider:
\begin{itemize}
\item $n_{l-1}=0$, $n_l\neq 0$

In this case
\begin{equation} \label{c1}
\bar{f}(0) f(n_l) - f(0)\bar{f}(n_l) = u(1, n_l\!-\!1 ) f(1) f(n_l\!-\!1) \;,
\end{equation}

\item $n_{l-1}\neq 0$, $n_l=0$

In this case
\begin{equation} \label{c2}
\bar{f}(n_{l-1}) f(0) - f(n_{l-1})\bar{f}(0) = - u(n_{l-1},0)
f(n_{l-1}) f(0) \;.
\end{equation}
The lhs of (\ref{c1}) and (\ref{c2}) are both of the same form,
therefore we are able to eliminate $\bar{f}(n)$ from these two
equations which yields
\begin{equation} \label{recursion}
f(n) = \frac{u(1,n-1)}{u(n,0)} \frac{f(1)}{f(0)} f(n-1)\;,
\end{equation} 
for $n>0$. This recursion is easily iterated to obtain the factors
$f(n)$ in terms of the hop rates:
\begin{equation} \label{mis f(n)}
f(n) = f(0) \left( \frac{f(1)}{f(0)}\right)^n\prod_{i=1}^n \left(
\frac{u(1,i-1)}{u(i,0)}\right)\;. 
\end{equation}
We can now substitute (\ref{recursion}) into (\ref{c1}) or
(\ref{c2}) to find the expression for $\bar{f}(n)$, hence
\begin{equation} \label{bar(f)(n)}
\bar{f}(n) = f(n) [\bar{f}(0) -u(n,0)]\;.
\end{equation}

\item $n_{l-1}\neq 0$, $n_l\neq 0$

We have used the first two cases to derive expressions for $f(n)$ and
$\bar{f}(n)$. Thus we must now satisfy (\ref{cancellation}) under
substitution of (\ref{recursion}) and (\ref{bar(f)(n)}). We find that
(\ref{cancellation}) is indeed satisfied provided the hop rates satisfy
the condition
\begin{equation} \label{mis constraint}
\fl u(n,m) - u(n+1,m-1) \frac{u(1,n) u(m,0)}{u(n+1,0) u(1,m-1)} = u(n,0) -
u(m,0)\;.
\end{equation}
This condition reduces to two separate conditions
\begin{eqnarray}
u(m,n) &=& u(n+1,m-1) \frac{u(1,n) u(m,0)}{u(n+1,0) u(1,m-1)} 
\label{db}\\
u(n,m)-u(m,n) &=& u(n,0)-u(m,0) 
\end{eqnarray}
for all values of $n$ and for $m>0$.

\end{itemize}
Hence the steady state of the Misanthrope process is given by the
factorised form (\ref{P(C)}) with $f(n)$ given by (\ref{mis f(n)})
provided the hop rates satisfy the constraint (\ref{mis constraint}).
We remark that if we take $u(n,m)$ to be a function of $n$, the number of
particles at the departure site only, we recover the ZRP, albeit via a
slightly different derivation to the one presented in section
\ref{definition}.

\subsubsection{Mapping to the KLS model}
\label{sec:KLS}
\begin{figure}
\begin{center}
\includegraphics[scale=0.6]{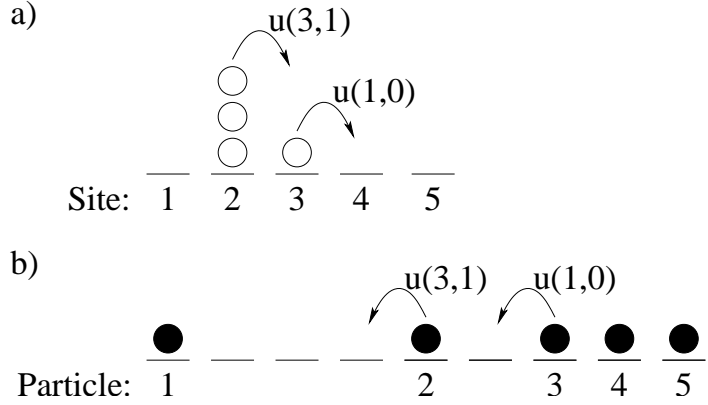}
\caption{Mapping between the Misanthrope process and the asymmetric
  exclusion process}
\label{fig:KLSmapping}
\end{center}
\end{figure}

Just as for the ZRP, the Misanthrope process can be mapped onto an
exclusion process. This mapping is illustrated in figure
\ref{fig:KLSmapping}. 
It is achieved in the same way as for the ZRP ---
sites in the Misanthrope process become occupied sites in the
exclusion process and particles in the Misanthrope process become
vacant sites in the exclusion process --- the only difference is that
now the hop rates depend not only on the distance to the next particle
in front, but also on the distance to the previous particle behind. 

An example of this mapping is between the Misanthrope process and the
KLS model \cite{KLS84}. The KLS model is an exclusion process in which
particles hop from site $i$ to site $i-1$ with a rate depending on the
occupancy of sites $i-2$ and $i+1$. The dynamics are defined by the
processes
\begin{eqnarray}
1 \, 0 \, 1 \, 1 \; \stackrel{\alpha_1}{\longrightarrow}  \; 1 \, 1 \,
0 \, 1 \;,\qquad \qquad 
1 \, 0 \, 1 \, 0 \; \stackrel{\alpha_2}{\longrightarrow}  \; 1 \, 1 \,
0 \, 0 \;,\nonumber \\ \label{KLSrates}
0 \, 0 \, 1 \, 0 \; \stackrel{\alpha_3}{\longrightarrow}  \; 0 \, 1 \,
0 \, 0 \;,\qquad \qquad 
0 \, 0 \, 1 \, 1 \; \stackrel{\alpha_4}{\longrightarrow}  \; 0 \, 1 \,
0 \, 1 \;,
\end{eqnarray}
where we use a `1' to represent a particle and a `0' to represent a
vacancy. The steady state probabilities $P(\{\eta_i\})$, where the
occupation variables $\eta_i=0$ or $1$, can be written in the form
\begin{equation} \label{KLS P(C)}
P(\{\eta_i\}) = \frac{1}{Z_{L,N}^{({\rm KLS})}} \;{\rm exp}\! \left( \beta
\sum_{i=1}^{L+N} \eta_i \eta_{i+1} \right) \;,
\end{equation}
for a system with $L$ particles and $N$ vacancies, where
$Z_{L,N}^{({\rm KLS})}$ is a normalisation. This steady state holds
provided \cite{KS90}
\begin{equation} \label{KLS constraint}
\alpha_2 = {\rm e}^\beta \alpha_4\;, \qquad
\alpha_2+\alpha_4=\alpha_1+\alpha_3\;.
\end{equation}
The mapping to the Misanthrope process is achieved by making the
identification: $u(n,m)=\alpha_3$ for $n>1$, $m>0$; $u(n,0)=\alpha_4$
for $n>1$; $u(1,m)=\alpha_2$ for $m>0$; $u(1,0)=\alpha_1$.  With this
choice of $u(n,m)$, condition (\ref{mis constraint}) reduces to
(\ref{KLS constraint}). This leads to a Misanthrope process for $N$
particles on $L$ sites with the steady state (\ref{P(C)}) where $f(n)$
is given by
\begin{equation}
f(n) = \left( \frac{f(1)}{f(0)} \right)^n f(0) \; {\rm e}^{\beta
(n-1)}\qquad \mbox{for}\quad n\ge 1 \;,
\end{equation}
and where substitution of the Misanthrope rates into (\ref{mis
constraint}) yields the second of the conditions in (\ref{KLS
constraint}). Thus, under the mapping the Ising-like form of the
steady state (\ref{KLS P(C)}) changes into a simple factorised form.

\subsubsection{Misanthrope  process on fully-connected geometry}
\label{sec:fcg}
In this subsection we consider a model where a particle can hop to any
other site with a rate which depends on the occupations of the
departure and destination sites.  Thus we are considering a
Misanthrope process on a fully-connected geometry.

Since the sites are fully connected the steady state weight should be
invariant under all permutations of the sites. Note, however that on a
finite system this does not imply a factorised steady state. This is
perhaps counterintuitive as one might expect the fully-connected
geometry to yield `mean-field' results and the steady state to
factorise.  On the other hand, in the limit of infinite system size
mean field results should be recovered and it would be interesting to
demonstrate that this is the case.

On a finite system, it turns out that the steady state has a
factorised form only if the hop rates satisfy a constraint.  This
constraint is less restrictive than the asymmetric one-dimensional
case discussed above. When this constraint is satisfied detailed
balance is obeyed. Thus, in this case the Misanthrope process is
equivalent to the urn models reviewed in \cite{GL02}.

To see this, consider the condition on the steady state probabilities: 
\begin{eqnarray} 
0 &= &\sum_{l\neq k}  \left[ u(n_l\!+\!1, n_k\!-\!1 )
  P(\cdots,n_{l}\!+\!1,\cdots, n_k\!-\!1, \cdots) 
-u(n_{l},n_k) P(\{n_l\})
 \right. \nonumber \\
&& + u(n_k\!+\!1, n_l\!-\!1 )
  P(\cdots,n_{l}\!-\!1,\cdots, n_k\!+\!1, \cdots)
   \left. 
  - u(n_{k},n_l) P(\{n_l\}) \right] \;.
\end{eqnarray}
where the sum runs over all pairs of sites $l,k$.
The assumption of a factorised steady state yields
\begin{eqnarray} 
0 &= &\sum_{l\neq k}  \left[ u(n_l\!+\!1, n_k\!-\!1 )
\frac{f(n_l\!+\!1)f(n_k\!-\!1 )}{f(n_l)f(n_k)}
-u(n_{l},n_k)
 \right. \nonumber \\
&&  \left.
+ u(n_k\!+\!1, n_l\!-\!1 )
\frac{f(n_k\!+\!1)f(n_l\!-\!1 )}{f(n_k)f(n_l)}  
  - u(n_{k},n_l) \right] \;.
\end{eqnarray}

It turns out that to satisfy this one must have detailed balance
\begin{equation} 
 u(n_l\!+\!1, n_k\!-\!1 )
f(n_l\!+\!1)f(n_k\!-\!1 )
= u(n_{k},n_l)f(n_l)f(n_k)
\end{equation}
which yields
\begin{equation}
f(n) = f(0) \left( \frac{f(1)}{f(0)}\right)^n
\prod_{m=1}^n  \frac{u(1,m-1)}{u(m,0)}
\end{equation}
provided the following condition holds for $n>1,m>0$
\begin{equation}
u(n,m)  = 
\frac{u(n,0)u(1,m)}{u(1,n-1)u(m+1,0)}
u(m+1,n-1)\;.
\label{Miscond}
\end{equation}
which is one of the two conditions (\ref{db}) required for the
 one-dimensional asymmetric case to factorise.  A simple way to
 guarantee the equality of (\ref{Miscond}) is for $u(n,m)$ to
 factorise into a function of $n$ and a function of $m$.  This is the
 case of interest for rewiring networks discussed in the next
 subsection.

\subsubsection{Mapping  to network dynamics}
\label{sec:rewire}
Recall that a network is a collection of nodes connected by edges
with dynamics for rewiring of edges (see section \ref{sec:net}).
Here we discuss how these dynamics correspond to Misanthrope dynamics.

Let us fix the number of nodes in the network to be $L$ and the number
of (undirected) edges to be $N$.  A general rewiring dynamics is as
follows \cite{DM03}: an edge is chosen at random and one end of the
edge initially connected to node $k$ is rewired to another node $k'$
with rate $\phi(k') h(k)$.  An increasing function $\phi(k')$
describes preferential attachment whereas a decreasing function $h(k)$
describes enhanced detachment of an edge from nodes with low degree.
(In the network literature $f(k)$ is usually used to denote $\phi(k)$
but we have chosen to use $\phi(k)$ to avoid a clash of notation with
the single site weights $f(n)$.)  Thus the overall rate of a rewiring
event such that two nodes with degrees $k,k'$ end up with degrees
$k-1,k'+1$ is proportional to $k \phi(k') h(k)$. Note that the factor
$k$ stems from choosing an edge at random. The dynamics clearly
preserves the number of edges and thus generates the `canonical
ensemble'.  Other ensembles of graphs are discussed in \cite{DMS03}
and \cite{BCK01}.
Note, however, that the dynamics may generate self-connected nodes
(tadpoles) and multiple edges between two nodes (melons). It has been
argued that these graphs have little effect on the resulting ensemble
\cite{DM03}.

We now think of the number of edges connected to a node as the number
of particles at a site in the Misanthrope process. The rate at which a
particle hops from a site with $k'$ to a site with $k$ particles is
then $u(k,k')= k h(k) \phi(k')$.  Also note that an edge may be
rewired from one node to any other, thus we are considering
Misanthrope dynamics on a fully-connected geometry. We may therefore
invoke the results of \ref{sec:fcg} which tell us that if $u(k,k')$
factorises, which is the case here, we have detailed
balance and the steady state factorises.  The
single-site weight becomes
\begin{equation}
f(k) = f(0) \left( \frac{f(1) h(1)}{f(0) \phi(0)}\right)^k
\prod_{m=1}^k \frac{\phi(m-1)}{m h(m)}
\end{equation}

Recalling the results of section \ref{sec:homcon} we see that for
condensation to occur we require that $m h(m)/\phi(m-1)$ decay
more slowly than $1+2/m$.  Setting $h(k)=1$ gives the preferential
attachment case.  Thus the preferential attachment function $\phi(m)/m$
should increase more quickly than $1-1/m$ to have condensation.

An alternative mapping from the dynamics of a  network  to a ZRP is 
proposed in \cite{PM04}.

\subsubsection{Relation to Backgammon model}
\label{sec:bgurn}

The Backgammon model, introduced in \cite{Rit95}, is an urn model
which can be related to the Misanthrope process.  There has been
considerable interest in the Backgammon model as a simple model of
glassy dynamics. Defined on a fully-connected geometry, the idea is
that particles hop between sites and there is an energy cost
associated with occupied sites. The hopping rate from site $l$ to $k$
is therefore a product of factors for each site and the existence of
an energy function guarantees that condition (\ref{db}) is satisfied.
Since there is an energy function, the dynamics can be parameterised
by a temperature and at very low temperatures the steady state will be
dominated by configurations with very few occupied sites. Glassy
dynamics are exhibited in the relaxation towards the steady state,
where the elimination of occupied boxes becomes slower and slower.

A simple example of Misanthrope dynamics corresponding to the
Backgammon model is $u(n,m) = 1+ ({\rm e}^{-\beta}-1) \delta_{m,0}$
for $n>0$, where $\beta$ is the inverse temperature.

\subsection{General mass transfer model}
\label{sec:EMZ}
Given that the steady state of the ZRP factorises it is natural to ask
under what conditions may a factorised steady state be admitted.  Recent
work \cite{EMZ04,ZEM04} has answered this question and revealed an
appealingly simple condition for factorisation within a framework that
encompasses a wide range of models.

The work \cite{EMZ04,ZEM04} generalises the ZRP in three ways
\begin{itemize}
\item The mass is a  continuous variable
\item Arbitrary amounts  of mass may move from one site to another
\item The dynamics comprises discrete (or parallel)
timesteps 
\end{itemize}

The class of models considered are defined on a one-dimensional
lattice of $L$ sites with periodic boundary conditions (site $L+1$=
site 1); associated with each site is a continuous mass variable
$m_i$, $ i=1\ldots L$. The total mass is given by $M=\sum_{i=1}^L
m_i$.

The dynamics is defined as follows: at each time step, at each site
$i$, mass $\mu _i$ drawn from a distribution $\phi (\mu _i|m_i)$
`chips off' the mass $m_i$, and moves to site $i+1$. Thus all sites
are updated in parallel and mass is transferred simultaneously between
sites.

\subsubsection{Steady state factorisation condition}
The central result of \cite{EMZ04} is a necessary and sufficient
 condition for the steady state to factorise which states that the
 chipping function must be of the form
\begin{equation}
\phi (\mu |m)=\frac{v\left( \mu \right) w\left( m-\mu \right) }{\left[
v*w\right] \left( m\right) }\;.
\label{phimum}
\end{equation}
In this case the single-site weights become
\begin{equation}
f(m) = \left[ v*w\right] \left( m\right)\;,
\label{fconv}
\end{equation}
where in expressions (\ref{phimum}) and (\ref{fconv}) the convolution
is defined as
\begin{equation}
\left[ v*w\right] \left( m\right)
\equiv \int_0^m d\mu \
v\left( \mu \right) w\left( m-\mu \right)   \label{fn}
\end{equation}
Equation (\ref{phimum}) expresses that the hopping function should be
a product of a function of $\mu$, the mass which moves, and a function
of $m-\mu$, the mass which remains, divided by a normalisation which
is a function of $m$.

The proof that (\ref{phimum}) is necessary and sufficient is presented
in \cite{EMZ04}. Here we content ourselves  with checking that it is a 
sufficient condition.

The steady state probabilities satisfy the following condition
\begin{equation}
\fl P(\underline{m})=\prod_{l=1}^L\int_0^\infty \ensuremath{\mathrm{d}}%
m_l^{\prime }\int_0^{m_l^{\prime }}\ensuremath{\mathrm{d}}\mu _l\ \phi (\mu
_l|m_i^{\prime })\prod_{k=1}^L\delta (m_k-m_k^{\prime }+\mu _k-\mu
_{k-1})\,P(\underline{m^{\prime }}) \;.
\label{Pgen}
\end{equation}
The r.h.s.\ of this equation is to be understood as integrating over
all possible configurations before the update: site $l$ contained mass
$m_l' = m_l+\mu_l-\mu_{l-1}$ and mass $\mu_l$ moved to site $l+1$ at
the update.  Assuming a factorised steady state, equation (\ref{Pgen})
reduces to
\begin{equation}
\fl \prod_{l=1}^L f(m_l)= \prod_{l=1}^L\left[ \int_0^{m_{l+1} }
\mathrm{d} \mu_l\ \right]
\prod_k \phi (\mu_k|m_k + \mu_k-\mu_{k-1}) f(m_k + \mu_k-\mu_{k-1}) \;. 
\end{equation}
Inserting (\ref{phimum},\ref{fconv}) the r.h.s becomes
\begin{equation}
\prod_{l=1}^L\left[ \int_0^{m_{l+1} }
\mathrm{d} \mu_l\ \right]
\prod_k v(\mu_k)w(m_k -\mu_{k-1})  
= 
\prod_{l=1}^L  \left[ v*w\right](m_l)\;,
\end{equation}
where we relabelled $k=l+1$.  Thus (\ref{Pgen}) is satisfied when
(\ref{phimum},\ref{fconv}) hold.

\subsubsection{Continuous time limit}
The continuous time case is a limit of the discrete time case when
we let the time step be ${\mathrm d}t$ and set $v(\mu) = \delta(\mu) +
x(\mu) {\mathrm d}t $ then take the limit ${\mathrm d}t \to 0$.
Condition (\ref{phimum}) reduces to requiring hopping {\em rates}
$\gamma(\mu|m)$ of the form
\begin{equation}
\gamma(\mu|m) = \frac{x(\mu )w(m-\mu )}{w(m)}
\label{gammum}
\end{equation}
in which case $f(m) = w(m)$.

It is useful to check explicitly that this works in a way similar to
section \ref{sec:ss}\;. The steady state probabilities satisfy the
following condition which generalises (\ref{ME})
\begin{eqnarray} 
\fl \lefteqn{0 = \sum_{l=1}^{L} \left[ 
\int_0^{m_{l}} {\mathrm d}\mu_{l-1}  \gamma(\mu_{l-1}|\mu_{l-1}+m_{l-1})
 P(\{\ldots, m_{l-1} + \mu_{l-1}, m_l - \mu_{l-1},\ldots \}) 
\right. }\nonumber &&
\\
&& \left. - \int_0^{m_l} {\mathrm d}\mu_l  \gamma(\mu_l|m_l)
 P(\{n_l\})  \right] \;.
\end{eqnarray}
Assuming a factorised steady state and equating terms $l$ in the sum, yields
\begin{eqnarray}
\lefteqn{\int_0^{m_{l}} {\mathrm d}\mu_{l-1}
  \gamma(\mu_{l-1}|\mu_{l-1}+m_{l-1}) f(m_{l-1} + \mu_{l-1})f(m_l - \mu_{l-1})
=}\nonumber \\
\int_0^{m_l} {\mathrm d}\mu_l  \gamma(\mu_l|m_l)
f(m_{l-1}) f(m_l)\;.
\end{eqnarray}
Inserting (\ref{gammum}) and $f(m) = w(m)$ yields the required
identity.

The case of discrete masses is easily obtained by restricting $\mu,
m-\mu$ to be positive integers in (\ref{phimum}) and defining the
convolution as
\begin{equation}
\left[ v*w\right] \left( m\right)
\equiv \sum_{\mu=0}^m 
v\left( \mu \right) w\left( m-\mu \right)\;.   
\end{equation}

\subsubsection{Heterogeneous case}
We also note that the condition (\ref{fn}) can be generalised to the
heterogeneous case where $\phi _l(\mu |m)$ depends on the site $l$.  A
necessary and sufficient condition for a factorised steady state is
that
\begin{equation}
\phi _l(\mu |m)=\frac{v\left( \mu \right) w_l\left( m-\mu \right) }{\left[
v*w_l\right] \left( m\right) }\;,
\end{equation}
where $v$ and $w_l$ are arbitrary functions but $v$ must be the same for
each site. The single-site weights  are given by 
\begin{equation}
f_l(m)=\left[ v*w_l\right] \left( m\right) \,\,.
\end{equation}

The continuous time limit implies that the site-dependent hopping
rates should be of the form
\begin{equation}
\gamma(\mu|m) = \frac{x(\mu )w_l(m-\mu )}{w_l(m)}\;,
\label{gammumh}
\end{equation}
in which case $f_l(m) = w_l(m)$.

\subsubsection{Test for factorisation}
Although (\ref{phimum}) is appealingly simple, it does not tell us
directly if a given hopping function $\phi$ yields a factorised steady
state. Happily, a direct test is easily constructed \cite{ZEM04}. We
require that
\begin{equation}
\left. \frac \partial {\partial \mu }\right| _\sigma \left. \frac \partial
{\partial \sigma }\right| _\mu \ln \phi \left( \mu |\mu +\sigma \right)
= h(\mu+\sigma)\;,
\label{test}
\end{equation}
i.e. the result of differentiating $\ln \phi$ with respect to $\mu$,
the mass that moves, and $\sigma$, the mass that remains, is a
function of $m =\sigma +\mu$ alone.  Then from (\ref{phimum}) we would
have $h(m) = - d^2 \ln f(m)/dm^2$ and integrating twice yields
\begin{eqnarray}
f\left( m\right) =\exp \left[ -\int^m \ensuremath{\mathrm{d}} m^{\prime
}\int^{m^{\prime }}\ensuremath{\mathrm{d}} m^{\prime \prime }h\left(
m^{\prime \prime }\right) \right] \,\,.  \label{fm-exp}
\end{eqnarray}

In the same way we can construct a test for the integer mass case:
we require 
\begin{equation}
\frac{\phi(\mu+1|n+2) \phi(\mu|n)}{\phi(\mu+1|n+1) \phi(\mu|n-1)}
=R(n)\;,
\label{testd}
\end{equation}
i.e. the cross-ratios on the l.h.s (defined when \emph{all} of the
$\phi $'s are positive) are functions of $n$ alone.  Then from
(\ref{phimum}) we would have
\begin{equation}
R(n)=\frac{f(n+1)^2}{f(n)\,f(n+2)}\;,
\end{equation}
which implies the recursion 
\begin{eqnarray}
\frac{f(n+2)}{f(n+1)}=\frac{1}{R(n)}\frac{f(n+1)}{f(n)}\;.  \label{recur1}
\end{eqnarray}
Iterating (\ref{recur1}) twice yields
\begin{equation}
f(n) = f(0)  \left( \frac{f(1)}{f(0)}\right)^n
\prod_{j=0}^{n-2}\left[ \prod_{k=0}^j \frac{1}{R(k)}\right]
\quad \mbox{for}\quad n\geq 2\;.  
\label{fnd}
\end{equation}

\subsubsection{ZRP with parallel dynamics}
As an illustration of the use of these tests and the construction of
factorised steady let us first consider the ZRP with parallel
dynamics.  At each time step a particle hops forward from a site with
$n$ particles with probability $u(n)$. Then $\phi(0|n) = 1- u(n)$,
$\phi(1|n) = u(n)$, $\phi(k|n) = 0$ for $k>1$.

The cross ratio in (\ref{testd}) is only defined for $\mu =0$, therefore
\begin{equation}
R(n) = \frac{u(n+2) u(n)}{u(n+1) u(n-1)}\;,
\end{equation}
is automatically a function of $n$ alone and factorisation is
guaranteed.  Expression (\ref{fnd}) becomes
\begin{equation}
f(n) = f(0) \left(\frac{f(1)}{f(0)}\right)^{n}  \frac{u(1)^n}{1-u(n)}
\prod_{j=1}^n \frac{1-u(j)}{u(j)}\;.
\end{equation}

This expression was first derived in the exclusion process context
\cite{Evans97} by a more complicated approach. Also in that work
expressions for $f(m)$ were presented for the case of ordered
sequential dynamics where one updates either in the forward sequence
$l = 1 \dots L$ or the backwards sequence $l= L \ldots 1$.

\subsubsection{ZRP with transfer of more than one mass unit}
As a more complicated example we consider a generalised ZRP where mass
`chunks' of size one or two can chip off at each time step with
probabilities $u_1(n)$ and $u_2(n)$ respectively. In this case we have
two cross ratios in (\ref{testd}) defined for $\mu =0,1$. Demanding
that these be equal implies
\begin{eqnarray}
R(n)&=& \frac{u_1\left( n+2\right) \left[ 1-u_1\left(
n\right) -u_2\left( n\right) \right] }{u_1\left( n+1\right) \left[
1-u_1\left( n+1\right) -u_2\left( n+1\right) \right] }
\label{Rgzrp}\\
&=&\frac{u_2\left( n+2\right) u_1\left( n\right) }{%
u_2\left( n+1\right) u_1\left( n+1\right) }
\quad\mbox{for}\quad n\geq 1 \;,
\label{R1}
\end{eqnarray}
which reduces to
\begin{equation}
\frac{u_2(n+1)(1-u_1(n)-u_2(n))}{u_1(n+1) u_1(n)} = A
\quad\mbox{for}\quad n\geq 1\;,
\label{Adef}
\end{equation}
where $A$ is a positive constant independent of $n$.  This recursion
fixes $u_2(n)$ in terms of $A, u_2(1)$ and $u_1(n)$.  Then we can use
(\ref{fnd}) and (\ref{Rgzrp}) to deduce the single-site weights
\begin{equation}
f(n) = f(0) \left(\frac{f(1)}{f(0)}\right)^{n}
 \frac{u_1(1)^n}{1-u_1(n)-u_2(n)}
\prod_{j=1}^n \frac{1-u_1(j)-u_2(j)}{u_1(j)}\;.
\end{equation}

\subsubsection{Condensation}
\label{sec:EMZcon}
We now analyse the condensation phenomena arising from $f(n)$ of the
form (\ref{fconv}).  For brevity we restrict our attention to the
homogeneous case.  The grand canonical partition function becomes
\begin{equation}
{\cal Z} (z)
= \left[ \int_0^\infty {\rm d}m\,  z^m \left[ v*w\right](m)\right]^L = 
 \left[ V(z)\, W(z)\right]^L\;,
\end{equation}
where
\begin{equation}
V(z) = \int_0^\infty {\rm d}\mu\, z^\mu v(\mu) \quad ; \quad
W(z) = \int_0^\infty {\rm d}\sigma\, z^\sigma w(\sigma) \; .
\end{equation}
As in (\ref{rhogc},\ref{rhocon}) the mass  density $\rho$ 
determines the fugacity $z$  through
\begin{equation}
\rho = 
 \frac{z V'(z)}{V(z)} +  \frac{zW'(z)}{W(z)}\;.
\label{phigce}
\end{equation}
As before the condensation mechanism is as follows: since
(\ref{phigce}) is an increasing function of $z$ one can always find a
finite solution of of equation (\ref{phigce}) provided the r.h.s.
diverges at the radius of convergence of $V$ or $W$.  On the other
hand, a condensation transition can occur if $zW'(z)/W(z)$ and $z
V'(z)/V(z)$ have finite limits as $z$ tends to the lower of the radii
of convergence of $V(z)$,$W(z)$.

Note the condensation mechanism may be driven by either or both of
$v(\mu)$, $w(\sigma)$ according to which of $V(z)$, $W(z)$ has a
finite radius of convergence.  The condensation in the ZRP corresponds
to condensation driven by $w(\sigma)$ since in the ZRP the amount of
mass that can chip off is restricted to one unit.  Condensation driven
by $v(\mu)$ would require a slowly decaying $v(\mu)$ which would
correspond to chipping off large fractions of mass. One would expect
that under these conditions the condensate is mobile since events
where a large fraction of mass is chipped off will move the condensate
around the lattice.

Thus in addition to the two types of condensation seen in the ZRP i.e.
condensation of particles onto the site with the slowest hopping rate
in a heterogeneous system and condensation due to slowly decaying hop
rates on a homogeneous system, we identify a third type of
condensation due to slowly decaying size distribution of the chunks of
mass transferred.

\subsection{ZRP with non-conservation}
\label{sec:nc}
So far we have considered the ZRP and generalisations where the
particle number, or mass, and site number is conserved.  In this
section we explore some cases where these quantities are allowed to
fluctuate yet the factorisation properties are preserved. These
dynamics allows one to generate different ensembles from the canonical
ensemble.  As we shall see, this admits a variety of new transitions.

We restrict ourselves to periodic boundary conditions although we note
that an open boundary ZRP has recently been considered in
\cite{LMS04}.

\subsubsection{Fluctuating particle number and generation of grand
  canonical ensemble}  
Let us consider dynamics where in addition to
the hopping rates $u(n)$, a particle is created at site $l$ with rate
$c(n_l)$ and a particle is annihilated with rate $a(n_l)$.  In order
for the steady state to remain factorised and for the particle number
to fluctuate, the dynamics for the creation and annihilation of
particles must obey detailed balance with respect to the steady state
\begin{equation}
P(\{ n_L \}) = \frac{\prod_l\left[ z^{n_l} f(n_l)\right]}{{\mathcal Z}}
\end{equation}
where $f(n)=f(n-1)/u(n)$ satisfies the usual factorisation condition
(\ref{rec}). Detailed balance with respect to particle non-conservation at
all  sites implies $a(n)f(n) z^n =  c(n-1)f(n-1)z^{n-1}$ so that
\begin{equation}
a(n) z = u(n) c(n-1)\;.
\end{equation}
A convenient solution of this equation is $a(n) =u(n)$ and $c(n) =
z$. Thus, with this choice of rates one generates in the steady state
the grand canonical ensemble of section \ref{sec:gce}.

If $c(n)/a(n) >1$ as $n \to \infty$, the particle number will diverge
and a steady state will no longer be attained.  In section
\ref{sec:gce} a cut-off on particle number, which implies $f(m)=0$ for
$m>N$, was introduced to circumvent this problem.  In order to
generate the cut-off dynamically one sets $f(N+1) = f(N)/\kappa$ where
$\kappa \to \infty$. Then one can take $u(N+1)= a(N+1) = \kappa \to
\infty$.

Finally, in section \ref{sec:gce} $z$ was taken as dependent on the total
particle number $z(N)$. This can be generated dynamically by letting the
creation rate be $c= z(N)$. Note that such a dependence of local particle
creation rate on the global number of particles in the system implies
non-local dynamics.

\subsubsection{Fluctuating site and particle  number}

\begin{figure}
\begin{center}
\includegraphics[scale=0.6]{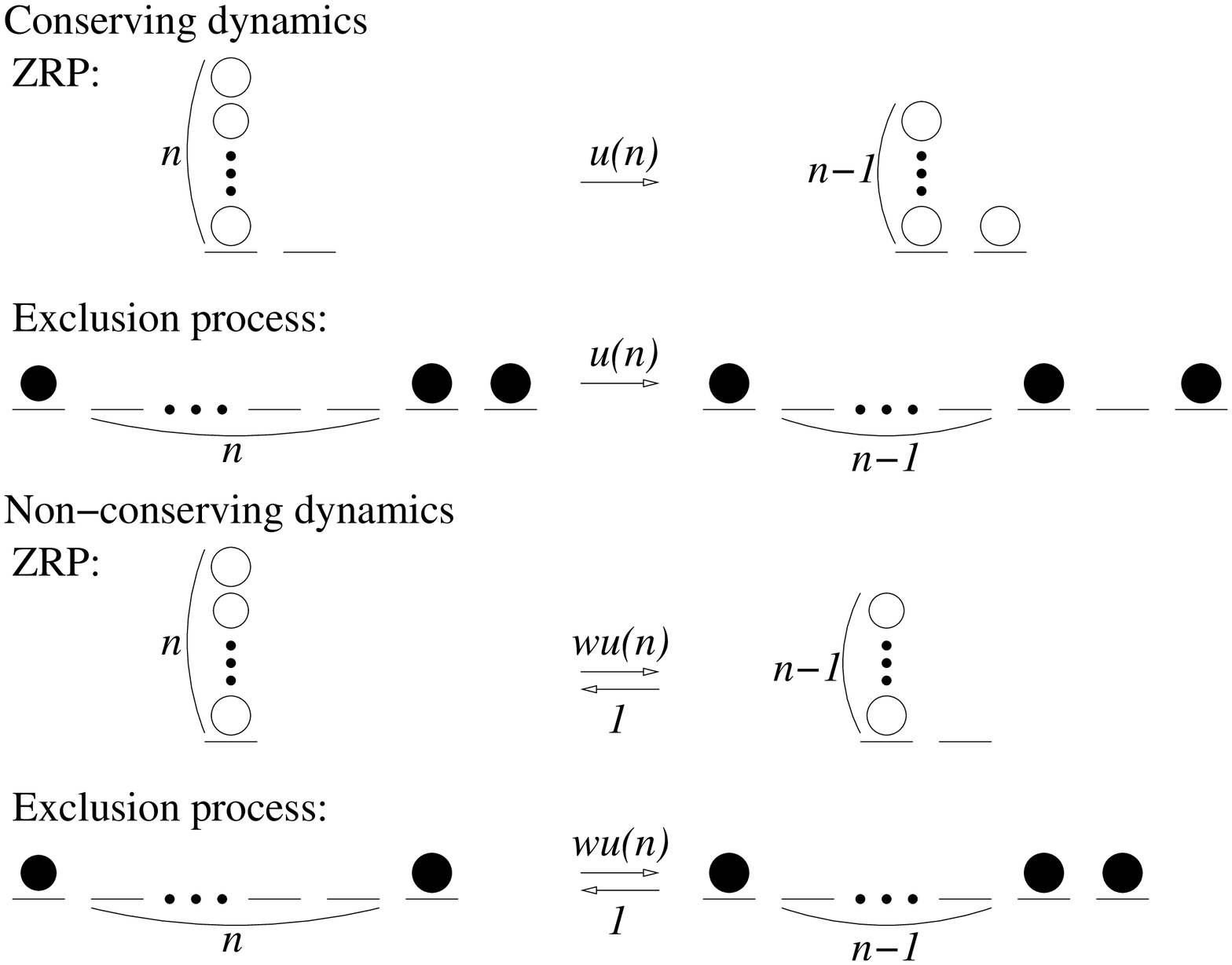}
\caption{Illustration of the dynamics which allows the site number and
the particle number to fluctuate with $L{+}N$ fixed. The conserving
dynamics, as shown at the top of the figure, are the usual ZRP
dynamics where particles hop to the neighbouring site with rate
$u(n)$.  The corresponding exclusion process dynamics are illustrated
beneath.  The lower part of the figure illustrates the non-conserving
dynamics: simultaneously a particle is annihilated and an adjacent
vacant site created with rate $w u(n)$ and the reverse process occurs
with rate one.  The corresponding exclusion process dynamics are shown
at the bottom of the figure.}
\label{fig:nonconmap}
\end{center}
\end{figure}

We now consider a microscopic dynamics which allows the number of
sites $L$ and the number of particles $N$ to fluctuate under the
constraint that the total number of sites plus particles remains
fixed: $L+N =K$. The dynamics is illustrated in figure
\ref{fig:nonconmap} along with the mapping to an exclusion process.

It is easy to check that this choice of creation and annihilation
dynamics satisfies detailed balance with respect to the steady state
\begin{equation}
P(\{ n_1 \ldots n_L \}) \propto
\prod_l \left[ w z^{n_l+1} f(n_l)\right]
\end{equation}
where, as usual, $f(n)=f(n-1)/u(n)$.  That is, the creation of new
sites balances annihilation of sites through
\begin{equation}
w u(n) \left[w z^{n+1} f(n)\right]=  \left[w z^n f(n-1)\right]\left[ w
  zf(0)\right]\;. 
\end{equation}

An interesting phase transition occurs under these dynamics which is
most easily studied in a grand canonical ensemble.  We define the
grand canonical partition function as
\begin{eqnarray}
{\mathcal Z}(z) &\equiv& \sum_{L=1}^\infty (w z)^L \prod_{l=1}^{L} 
\sum_{n_l=0}^\infty  z^{n_l}f(n_l)\\
&=& \frac{wzF(z)}{1-wzF(z)}
\end{eqnarray}
where $F(z)$ is defined as in (\ref{Fdef}). The value of $z$ is fixed
through the condition
\begin{equation}
K= z \frac{\partial \ln  {\mathcal Z}}{\partial z} =
\langle L \rangle \left[ \langle n \rangle  +1 \right]
\label{Kcon}
\end{equation}
where 
\begin{eqnarray}
\langle L \rangle = w \frac{\partial \ln  {\mathcal Z}}{\partial w}
= \frac{1}{1-wzF(z)}\label{La}\\
\langle n \rangle = 
\frac{z F'(z)}{F(z)}\;.
\label{nav}
\end{eqnarray}
$\langle L \rangle$ is the average number of sites and $\langle n
\rangle$ is the average number of particles per site.  When $K$ is
large, (\ref{Kcon}) is satisfied when either $\langle L \rangle$ or
$\langle n \rangle$ is large.

Let us take as example the solvable case of section \ref{sec:sc}:
$u(n) = 1 + b/n$ and for simplicity we consider only $b$ noninteger.

If $b <1$, $F(1)$ diverges and we can clearly find $z<1$ such that
$1-wzF(z) = {\mathcal O}(1/K)$.  Thus for $b<1$ we have from
(\ref{La},\ref{nav}) that $\langle L \rangle = {\mathcal O}(K)$ and
$\langle n \rangle = {\mathcal O}(1)$. This implies a number of sites
of order $K$ each with a finite number of particles.

For $b>1$ we have the explicit expression $F(1)= b/(b-1)$. Thus if $w
> w_c = (b-1)/b$ we can again choose $z$ to give $\langle L \rangle =
{\mathcal O}(K)$ and $\langle n \rangle = {\mathcal O}(1)$. For
$w<w_c$, on the other hand, we have different behaviour according to
whether $b < 2$ or $b>2$

For $b<2$, $F'(1)$ diverges. Therefore for $w<w_c$ we can choose
$z \nearrow 1$ such that $\langle n \rangle = {\mathcal O}(K)$.
In  this case $\langle L \rangle = {\mathcal O}(1)$, thus we have a finite
number of sites each with a number of particles of order $K$.

For $b>2$, $F'(1)= b/(b-1)(b-2)$, and for $w<w_c$ we cannot satisfy
(\ref{Kcon}) as $z \nearrow 1$.  Moreover as $z \nearrow 1$, both
$\langle L \rangle$ and $\langle n \rangle$ are finite.  Therefore we
must have a condensation transition wherein a condensate site emerges
which contains fraction one of the particles in the large $K$ limit.

The mechanism for these transitions was first studied in a simple
lattice model for DNA denaturation \cite{PS}.  Within the mapping to
an exclusion process the constraint that $K$ is constant corresponds
to fixed number of lattice sites with particle non-conservation.  Thus
the transition is from a state with a finite density of particle to a
zero density state which comprises a finite number of particles with
large gaps between for $1<b<2$, or the condensed state with one large
gap and a small jam of closely spaced particles for $b>2$.

It has been shown how the transition may occur in an exclusion process
with local dynamics by using two species of particles to generate {\em
effective} hop rates such that $b=3/2$ and $f(n) \sim 1/n^{3/2}$, in
the spirit of section \ref{sec:ps} \cite{EKLM02}.

\subsubsection{Fluctuating site number only}

One can also consider the case where only the site number $L$ is
allowed to fluctuate \cite{BBBJ}.  For dynamics in which, in addition
to the usual ZRP hop rates, vacant sites can be created with rate $w$
and annihilated with rate $1$, the steady state
\begin{equation}
P_L(\{ n_1 \ldots n_L \}) \propto w^L
\prod_{l=1}^L f(n_l)\;,
\end{equation}
satisfies detailed balance with respect to the creation and
annihilation dynamics. The $L$ superscript here is intended to
emphasise that this is the probability for a configuration of $N$
particles on exactly $L$ sites.

This model undergoes a transition which can be analysed in a way
following just the same steps as the previous case, so we neglect the
details and quote only the results. We consider hop rates
$u(n)=1+b/n$, and we consider the three quantities $N$, $\langle L
\rangle$ and $\langle n \rangle$.

There exists some critical value of $w$, $w_c$, such that if $w>w_c$
then we can work in the grand canonical ensemble and find a fugacity
such that $\langle L \rangle = {\mathcal O}(N)$ and $\langle n \rangle
= {\mathcal O}(1)$. However for $w<w_c$ two cases ensue. If $b<2$, one
can find a fugacity such that $\langle n \rangle = {\mathcal O}(N)$
and $\langle L \rangle = {\mathcal O}(1)$. On the other hand, if $b>2$
one finds that both $\langle L \rangle = {\mathcal O}(1)$ and $\langle
n \rangle = {\mathcal O}(1)$. Therefore in the limit $N\to \infty$, a
single site contains a fraction equal to one of the $N$ particles in
the system.


\section{Summary and open questions}
\label{summary}
In this work we have reviewed the properties of the zero-range process
and a variety of generalisations and related models.  We have focussed
on the property of a factorised steady state which is exhibited in
many of these models: the ZRP always factorises, whereas
generalisations such as the two-species model (section
\ref{sec:2szrp}) and Misanthrope process (section \ref{sec:Mis}),
where the hop rate also depends on the target site, only factorise
when the hop rates satisfy certain specific conditions
(\ref{CE},\ref{mis constraint}).  The generalisation of section
\ref{sec:EMZ}, to continuous mass, discrete time and arbitrary amount
of mass transferred, yields a condition for factorisation with an
appealingly simple general form.  The factorisation property allows an
exact analysis of steady state behaviour, in particular
condensation. It should be noted that condensation is not restricted
to models with factorised steady states, rather, factorisation, has
offered, so far, the only opportunity to study condensation exactly.
We remark that factorisation has also been used as a mean-field-type
approximation to study condensation in systems where the steady state
is not known exactly.

More generally, it would be of interest to study the structure of
non-factorised steady states and how condensation arises.  For
example, one generalisation of a factorised steady state is to a
matrix product state.

Returning to the case of factorised steady state, there have been a
number of developments in our understanding of condensation.  Here we
have reviewed the distinct mechanisms of condensation in homogeneous
systems (sections \ref{homogeneous}, \ref{sec:EMZcon}), heterogeneous
systems \ref{sec:hetero}, including the case of a single defect site
\ref{sec:sds}, non-conserving systems \ref{sec:nc} and systems with
more than one species of particle \ref{sec:2szrp}. We have also seen
an example where the condensate is mobile rather than fixed
\ref{sec:EMZcon}. In particular the analysis of condensation within
the canonical ensemble \ref{sec:homcon} should allow a deeper
understanding of the nature of condensates.

We have also reviewed the dynamics of condensation. The coarsening
dynamics is now typically understood at a mean-field level or using
heuristic random walk arguments. These arguments have been applied to
condensation in homogeneous \ref{sec:homdyn}, heterogeneous
\ref{sec:het dyn} and two-species systems \ref{sec:2sdyn}.  An open
question is whether one can obtain scaling functions for dynamics
beyond the mean field level.  This would enable one to address
questions of universality. We also note that for certain parameters
(for which condensation does not occur) the Bethe ansatz has been used
to calculate exactly dynamical properties of the ZRP \cite{P04}.

Finally, we are aware that this review is of a theoretical nature and
we have not emphasized or speculated upon experimental
realisations. In section \ref{sec:sgg} we discussed clustering in
shaken granular gases, which furnishes a pleasing experimental example
of condensation in interacting particle systems.  Another 
realisation of the ZRP noted in \cite{BM04} concerns the exchange of monomers
between protein filaments. It remains an important challenge
to establish further experimental instances of interacting particle
systems and the associated phenomena reviewed here.

\ack We would like to thank the following colleagues with whom it has
been a pleasure to discuss or to collaborate on the topics reviewed
here: Andrew Angel, Mustansir Barma, Richard Blythe, Mike Cates,
Bernard Derrida, Claude Godr\`eche, Stefan Grosskinsky, Kavita Jain,
Yariv Kafri, Joachim Krug, Erel Levine, Satya Majumdar, PK Mohanty,
David Mukamel, Owen O'Loan, Andreas Schadschneider, Gunter Sch\"utz,
Janos T\"or\"ok and Royce Zia.


\end{document}